\documentclass[pra,reprint]{revtex4-1}
\usepackage{graphicx}
\usepackage{appendix}
\usepackage{braket}
\usepackage{rotating}
\usepackage{simplewick}
\usepackage{amssymb}
\usepackage{amsmath}
\usepackage{txfonts}
\usepackage{bm}
\usepackage{color}
\usepackage{multirow}
\usepackage{booktabs}
\usepackage{dcolumn}

\begin{document}

\title{Low-temperature breakdown of many-body perturbation theory for thermodynamics}

\author{So Hirata}
\email{sohirata@illinois.edu}
\affiliation{Department of Chemistry, University of Illinois at Urbana-Champaign, Urbana, Illinois 61801, USA}

\date{\today}

\begin{abstract}
It is shown analytically and numerically that the finite-temperature many-body perturbation theory
in the grand canonical ensemble has zero radius of convergence at zero temperature  
when the energy ordering or degree of degeneracy for the ground state changes with the perturbation strength.
When the degeneracy of the reference state is partially or fully lifted at the first-order Hirschfelder--Certain degenerate
perturbation theory, the grand potential and internal energy diverge as $T \to 0$.
Contrary to earlier suggestions of renormalizability by the 
chemical potential $\mu$, this nonconvergence, first suspected by W. Kohn and J. M. Luttinger, 
is caused by the nonanalytic nature of 
the Boltzmann factor $e^{-E/k_\text{B}T}$ at $T=0$, also plaguing the canonical ensemble, which does not involve $\mu$. 
The finding reveals a fundamental flaw in perturbation theory, which is deeply rooted in 
the mathematical limitation of power-series expansions and is unlikely to be removed
within its framework.
\end{abstract}

\maketitle 

\section{Introduction}

In 1960, Kohn and Luttinger \cite{kohn} pointed out a possible mathematical inconsistency between the finite-temperature 
perturbation theory \cite{bloch,balian,blochbook,thouless1972quantum,mattuck1992guide,march1995many,Fetter,SANTRA} and its zero-temperature counterpart \cite{moller,Hirschfelder,szabo,shavitt}: The second-order grand potential $\Omega^{(2)}$ in the zero-temperature limit and 
second-order energy $E^{(2)}$ of many-body perturbation theory (MBPT) \cite{moller,szabo,shavitt} 
can differ from each other by divergent ``anomalous'' contributions for a degenerate, nonisotropic reference wave function. 
On this basis, they concluded that ``the BG [Brueckner--Goldstone perturbation]
series is therefore in general not correct'' \cite{kohn}.
For isotropic systems 
such as a homogeneous electron gas (HEG), the same authors showed that 
the difference is exactly compensated for by the terms containing the chemical potential $\mu$. This partial solution was generalized by Luttinger and Ward \cite{luttingerward} and by Balian, Bloch, and De Dominicis \cite{balian}.

The question posed by Kohn and Luttinger \cite{kohn} and the partial solution for isotropic systems 
may, however, be challenged in the following three respects:\ First, $\Omega$ and $E$ are separate
thermodynamic functions and are not expected to agree with each other at $T=0$; instead, the internal energy $U$ at $T=0$ should be more rigorously compared
with $E$. Second, such perturbation correction formulas for $U$ were unknown until recently \cite{HirataJha,HirataJha2} since the finite-temperature perturbation theory of Bloch and coworkers \cite{bloch,balian,blochbook} (see also Refs.\ \onlinecite{thouless1972quantum,mattuck1992guide,march1995many,Fetter,SANTRA}) adopts an
unequal treatment \cite{JhaHirata} of $\Omega$, $U$, and $\mu$.
Third, $E^{(2)}$ of MBPT may be already divergent
in a degenerate, extended system such as a HEG, obscuring the comparison; for a degenerate reference, $E^{(2)}$ from
the Hirschfelder--Certain degenerate perturbation theory (HCPT) \cite{Hirschfelder} should be used as the correct
zero-temperature limit, which is always finite for a finite-sized system.

In short, the Kohn--Luttinger conundrum remains to be an open question, implying that the finite-temperature perturbation theory may still be incorrect in a general sense, in particular, for a degenerate, nonisotropic 
reference wave function.

Recently, we introduced \cite{HirataJha,HirataJha2} a finite-temperature perturbation theory for electrons in the grand canonical ensemble wherein $\Omega$, $U$, and $\mu$ are expanded in power series on an equal footing. Two types of analytical formulas
were obtained for up to the second order in a time-independent, algebraic (nondiagrammatic) 
derivation:\ sum-over-states (SoS) and sum-over-orbitals (reduced) formulas. They reproduce numerically 
exactly the correct benchmark data \cite{JhaHirata} obtained as the $\lambda$-derivatives of the corresponding 
thermodynamic functions evaluated by the thermal full-configuration-interaction (FCI) method \cite{Kou} 
with a perturbation-scaled Hamiltonian $\hat{H} = \hat{H}_0 + \lambda \hat{V}$.
They permit a rigorous comparison of  
the zero-temperature limit of $U^{(n)}$ against $E^{(n)}$ of HCPT both analytically
and numerically. 
We can repeat this comparison for the finite-temperature perturbation theory in the canonical ensemble,
whose SoS formulas for the Helmholtz energy ($F$) and internal energy ($U$) have been reported up to the third order \cite{JhaHirata_canonical}.

In what follows, we will show analytically and numerically that for an ideal gas of identical 
molecules with a degenerate ground state,
 $U^{(1)}$ converges at a finite, but wrong zero-temperature limit. 
 For the same system, 
the zero-temperature limit of $U^{(2)}$ is divergent and clearly wrong since the correct zero-temperature limit ($E^{(2)}$ of HCPT) is always finite. 
While the chemical potentials 
$\mu^{(n)}$ ($0 \leq n \leq 2$) converge at the correct zero-temperature limits in our example, 
the grand potentials $\Omega^{(n)}$ ($1 \leq n \leq 2$) display the same nonconvergent (or even divergent) behaviors as $U^{(n)}$.  
Taken together, these findings justify the original concern of Kohn and Luttinger \cite{kohn} and establish
that the finite-temperature perturbation theory in the grand canonical ensemble 
is indeed incorrect in a general sense: Beyond the zeroth-order Fermi--Dirac theory, 
the perturbation theory for $U$ and $\Omega$ has zero radius of convergence at $T=0$ and 
becomes increasingly inaccurate at lower temperatures whenever the reference wave function
differs qualitatively from the true ground-state wave function.

The root cause of the failure does not have so much to do with the chemical potential $\mu$ (as implied by 
other authors \cite{kohn,luttingerward,balian})\ as with 
the smooth nonanalytic nature of the Boltzmann factor $e^{-E/k_\text{B}T}$ at $T=0$. 
The nonconvergence, therefore, persists in the canonical ensemble also \cite{JhaHirata_canonical}, which does not involve $\mu$.
It reveals the fundamental limitation of perturbation theory for thermodynamics, reminiscent 
of similar divergences in quantum electrodynamics \cite{sakurai,weinberg,dyson_physicsworld}.

\section{Illustrations}

Before going into the analytical formulas of $U^{(n)}$, $\mu^{(n)}$, and $\Omega^{(n)}$ and their numerical behavior for a molecular gas in Sections \ref{sec:KLtest}--\ref{sec:mu}, 
we will use three simple models to illustrate the essence of the breakdown of the thermodynamic perturbation theory. Nonconvergence 
is caused by the nonanalytic nature of $U$ at $T=0$ for a degenerate or qualitatively wrong reference (zeroth-order) wave function, 
preventing $U$ from being expanded in a converging power series.
This, in turn, originates from the nonanalytic nature of $e^{-E/k_\text{B}T}$ at $T=0$. This problem is unseen in the zero-temperature perturbation theory \cite{Hirschfelder} or variational finite-temperature  theory \cite{Kou}, but may be reminiscent of the theory of superconductivity whose
interaction operator has a similar form, $\delta e^{-1/\rho v}$ \footnote{``It should be noted that, although the latter  
is very small, the functional form of it is such that it cannot be expanded in a power series in the interaction parameter $v$, and thus in any many-body generalization
of the above method, perturbation theory would not be easy to apply.''\ (p.224 of Ref.\ \onlinecite{march1995many}).}.
This is, therefore, a manifestation of a fundamental mathematical limitation in the power-series expansions of
pathological functions and may be hard to resolve (e.g., by renormalization) within the framework of perturbation theory. 

\begin{figure*}
\includegraphics[width=0.95\textwidth]{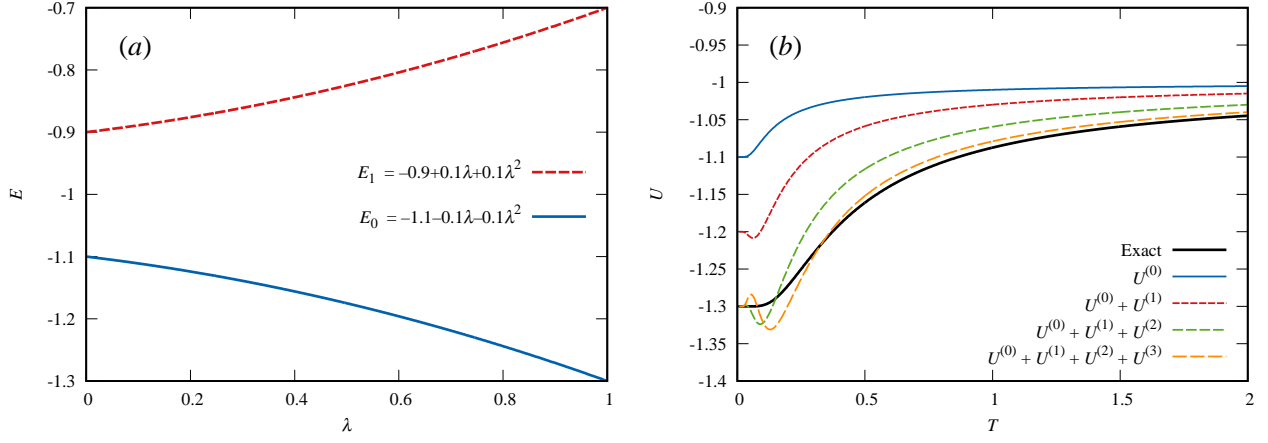}
\caption{({\it a}) $E_0$ and $E_1$ as a function of $\lambda$. ({\it b}) $U$ as a function of $T$ at $\lambda=1$ and its Taylor-series approximations, simulating an everywhere convergent perturbation expansion for a nondegenerate, correct reference. \label{fig:1}}
\end{figure*}

\begin{figure*}
\includegraphics[width=0.95\textwidth]{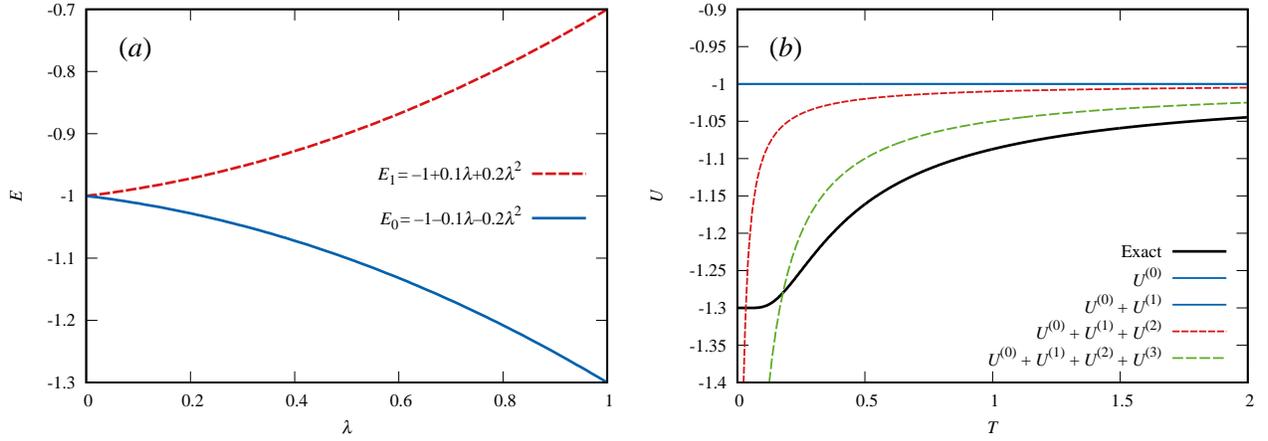}
\caption{({\it a}) $E_0$ and $E_1$ as a function of $\lambda$. ({\it b}) $U$ as a function of $T$ at $\lambda=1$ and its Taylor-series approximations, simulating a divergent perturbation expansion at $T=0$ for a degenerate reference. \label{fig:2}}
\end{figure*}

\begin{figure*}
\includegraphics[width=0.95\textwidth]{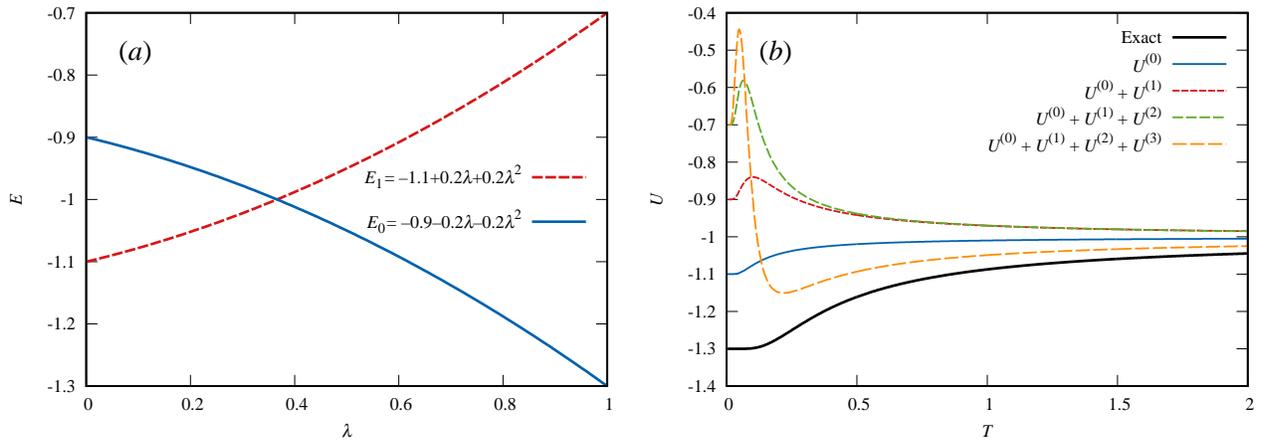}
\caption{({\it a}) $E_0$ and $E_1$ as a function of $\lambda$. ({\it b}) $U$ as a function of $T$ at $\lambda=1$ and its Taylor-series approximations, simulating a perturbation expansion convergent at a wrong limit at $T=0$ for an incorrect reference. \label{fig:3}}
\end{figure*}

Let us consider a function $U(T)$, which is an exponential-weighted average of $E(\lambda)$:
\begin{eqnarray}
U(T) = \frac{E_0(\lambda) e^{-E_0(\lambda)/T} + E_1(\lambda) e^{-E_1(\lambda)/T}}{e^{-E_0(\lambda)/T} + e^{-E_1(\lambda)/T}}.
\end{eqnarray}
This function is meant to capture the essential mathematical features of the internal energy (thermal average of energy) $U$ as a function of temperature $T$ in the canonical ensemble of a two-state system with energies $E_0(\lambda)$ and $E_1(\lambda)$. 
These energies are, in turn, functions of $\lambda$ (the perturbation strength), which simulate how they evolve from the zeroth-order reference ($\lambda=0$) to
the fully interacting limit ($\lambda=1$) of the system described by the Hamiltonian $\hat{H} = \hat{H}_0 + \lambda \hat{V}$. 

In Figs.\ \ref{fig:1}--\ref{fig:3}, we plot $U$  at $\lambda=1$ as a function 
of $T$ and its truncated Taylor-series approximations in $\lambda$ for three different sets of  $E_0(\lambda)$ and $E_1(\lambda)$  (which are also included in the respective figures). In all cases, $E_0(1)$ and $E_1(1)$ are always 
equal to $-1.3$ and $-0.7$, respectively, and, therefore, the exact $U$ (the thick solid black curves) in the fully interacting limit ($\lambda=1$) 
have the identical form, which is infinitely differentiable everywhere. 

Figure \ref{fig:1} shows that, when $E_0(\lambda)$ and $E_1(\lambda)$ do not cross or touch each other 
in the domain $0 \leq \lambda \leq 1$, 
the Taylor-series expansion of $U$ in $\lambda$ is finite and convergent everywhere at the correct limit. 
In the physics context, this corresponds to the case where
the perturbation theory for the internal energy $U$ is valid at all temperatures
and converges at the correct zero-temperature limit, $E_0(1)$,
when the reference chosen is nondegenerate and correct. By ``correct,'' we mean that the energy  ordering 
of the ground and excited states is unchanged in $0 \leq \lambda \leq 1$, with the zeroth-order ground state,
$E_0(0)$, smoothly morphing into (without crossing) the true ground state 
in the fully interacting limit, $E_0(1)$.

Figure \ref{fig:2} considers the case in which the internal energy $U$ in the canonical ensemble 
is expanded in a perturbation series with a degenerate reference. Here, a ``degenerate'' reference means 
that the degree of degeneracy of the true ground state, $E_0(\lambda)$, is partially or fully lifted as $\lambda=0 \to 1$.
It can be seen that the zeroth- and first-order Taylor-series approximations are constant, but the second-
and all higher-order approximations are divergent at $T=0$; the radius of convergence of the Taylor series of $U$ 
is zero at $T=0$. To paraphrase, when $E_0(0) = E_1(0)$, 
$U$ becomes a  nonanalytic function of $\lambda$ at $T=0$, 
which is infinitely differentiable yet not expandable in a converging power series. 

In Fig.\ \ref{fig:3}, we consider the third scenario, in which $U$ is expanded 
from a qualitatively wrong reference: 
The ground state in the zeroth-order description ($\lambda=0$) and thus the reference state evolves 
into the first excited state in the fully interacting limit ($\lambda=1$) and vice versa.
The Taylor-series approximations remain finite at any $T$, but 
converge at $E_1(1)=-0.7$ at $T=0$ (at the second and higher orders) 
instead of the correct zero-temperature limit of $E_0(1)=-1.3$.
Therefore, the perturbation theory for the internal energy $U$ becomes increasingly inaccurate at low $T$ and fails 
to converge at the correct zero-temperature limit when the reference is qualitatively wrong and does not
smoothly transform into the true ground state as $\lambda = 0 \to 1$. This is closely related to 
quantum phase transitions at $T=0$ caused by a modulation of the Hamiltonian \cite{QPT}, in this case, $\lambda$.

\section{The Kohn--Luttinger tests\label{sec:KLtest}} 

The internal energy $U$ in the grand canonical ensemble of electrons is the thermal average of energy,
\begin{eqnarray}
U = \frac{\sum_I E_I e^{-\beta E_I + \beta \mu N_I}}{\sum_I e^{-\beta E_I + \beta \mu N_I}}, \label{U_FCI}
\end{eqnarray}
where $I$ runs over all states with any number of electrons, $\beta=(k_\text{B}T)^{-1}$, $\mu$ is the chemical potential, 
and $E_I$ and $N_I$ are the exact (FCI) energy and number of electrons in the $I$th state, respectively.

A perturbation expansion of $U$ means
\begin{eqnarray}
U = U^{(0)} + \lambda U^{(1)} + \lambda^2 U^{(2)} + \lambda^3 U^{(3)} + \dots,
\end{eqnarray}
or, equivalently,
\begin{eqnarray}
U^{(n)} = \left.\frac{1}{n!} \frac{\partial^n U(\lambda)}{\partial \lambda^n}\right|_{\lambda=0},  \label{U_lambda}
\end{eqnarray}
where $U(\lambda)$ is given by Eq.\ (\ref{U_FCI}) whose $E_I(\lambda)$ is the $I$th eigenvalue (FCI energy) of a perturbation-scaled
Hamiltonian, $\hat{H}_0 + \lambda \hat{V}$. 
The corresponding perturbation expansion of $E_I$ is given by
\begin{eqnarray}
E_I^{(n)} = \left.\frac{1}{n!} \frac{\partial^n E_I(\lambda)}{\partial \lambda^n}\right|_{\lambda=0},  \label{E_lambda}
\end{eqnarray}
which is identified \cite{HirataJha,HirataJha2} as the $n$th-order HCPT correction \cite{Hirschfelder} to the $I$th-state energy, distinguished from the M{\o}ller--Plesset perturbation theory (MPPT) \cite{moller,szabo,shavitt} 
when the reference is degenerate.
Since many zeroth-order (excited, ionized, etc.)\ states are degenerate, it is imperative to use the 
degenerate perturbation theory that computes energy corrections that match the above definition
and remain finite for any state in a finite-sized system.
In contrast, a nondegenerate
perturbation theory such as MPPT diverges for a 
degenerate reference and is, therefore, inappropriate here, although HCPT reduces to MPPT for a nondegenerate reference. 
In this article, the acronyms MPPT, MBPT, and diagrammatic BG perturbation theories are used interchangeably,
but in distinction to HCPT.

The zero-temperature limit of $U$ is $E_0$ (the FCI energy for the true ground state) according to Eq.\ (\ref{U_FCI}), where 
the states are numbered in the ascending order of the FCI energy. 
Then, the correct zero-temperature limit of $U^{(n)}$ should be $E_0^{(n)}$, the latter being defined by HCPT for the true 
ground state, i.e., the lowest-energy state of the neutral molecule according to FCI. 
We, therefore, begin by generalizing the question
raised by Kohn and Luttinger \cite{kohn}: We ask whether the identity,
\begin{eqnarray}
\lim_{T \to 0} U^{(n)} \stackrel{?}{=} E_0^{(n)}\,\,\,(\text{the first KL test}), \label{KL_test}
\end{eqnarray}
holds in an ideal gas of identical molecules with a degenerate or nondegenerate reference, 
where $E_0^{(n)}$ is the $n$th-order HCPT energy correction for the lowest-lying neutral state of the molecule 
as per FCI.
We call this the first Kohn--Luttinger (KL) test.

The revised question eliminates many of the confusions sown by the original one. First, we are no longer comparing the zero-temperature limit of $\Omega^{(n)}$ with $E_0^{(n)}$, which differ from each other 
by a nonvanishing term involving $\mu^{(n)}$ \cite{JhaHirata}.
Second, $E_0^{(n)}$ is identified as the $n$th-order HCPT energy correction, and not as the
$n$th-order MPPT energy correction, the latter being ill-defined for a degenerate reference. 
Third, we apply the perturbation theory to an ideal gas of general molecules with a degenerate or nondegenerate reference (whose $E_0^{(n)}$ and $E_0$ are always finite) instead of 
a less general and problematic case of HEG, whose $E_0^{(2)}$ is
divergent for a multitude of reasons \cite{Hirata_KL}. 

We will also consider the second KL test which examines if $\mu^{(n)}$ converges
at the correct zero-temperature limit:
\begin{eqnarray}
\lim_{T \to 0} \mu^{(n)} \stackrel{?}{=} \frac{E^{(n)}_\text{anion} - E^{(n)}_\text{cation}}{2} \,\,\,(\text{the second KL test}), \label{KL_test2}
\end{eqnarray}
where $E^{(n)}_\text{anion}$ and $E^{(n)}_\text{cation}$ are the $n$th-order HCPT energy corrections for the anion and cation ground states, respectively. A justification for the right-hand side as the correct zero-temperature limit is given in Appendix \ref{app:mu_limit}.

The grand potential $\Omega$ bears the following relationship with
$U$, $\mu$, and entropy $S$:
\begin{eqnarray}
\Omega = U - TS -\mu \bar{N},
\end{eqnarray}
where $\bar{N}$ is the average number of electrons that keeps the system electrically neutral \cite{JhaHirata,HirataJha,HirataJha2}.
Differentiating this equation with respect to $\lambda$ and taking the $T \to 0$ limit, we arrive at the third KL test,
\begin{eqnarray}
\lim_{T \to 0} \Omega^{(n)} \stackrel{?}{=} E_0^{(n)} - \lim_{T \to 0} \mu^{(n)}\bar{N} \,\,\,(\text{the third KL test}), \label{KL_test3}
\end{eqnarray}
which is the closest to the original question posed by Kohn and Luttinger \cite{kohn} except that their chemical potential $\mu$ 
was determined variationally, further complicating the issue. With the analytical formulas for $\mu^{(n)}$ \cite{HirataJha,HirataJha2}, this test is equivalent to the union of the first two tests, and will 
be discussed only briefly in relation to the ``anomalous'' diagrams of Kohn and Luttinger \cite{kohn}.

In Sec.\ \ref{sec:U}, we will apply the first KL test to the SoS analytical formulas of $U^{(n)}$ ($0 \leq n \leq 2$)
in the grand canonical ensemble. Since the SoS and reduced analytical formulas are mathematically equivalent, 
they display the identical $T \to 0$ behaviors, leading to the same conclusion. We will, therefore, relegate the discussion of 
the reduced analytical formulas of $U^{(n)}$  to Appendix \ref{app:U}.
We then elucidate the zero-temperature limits of $\mu^{(n)}$ in Sec.\ \ref{sec:mu} using their reduced analytical formulas 
to see if they pass the second KL test. 
We then analyze the $T \to 0$ behaviors of $\Omega^{(n)}$ using their reduced analytical formulas
in relation to the anomalous diagrams in Appendix \ref{app:Omega}. 
In each section, we demonstrate the correctness of the analyses
by a numerical example of the square-planar H$_4$ molecule, which has a degenerate and incorrect reference.
Owing to the isomorphism of the SoS analytical formulas between the grand canonical ensemble \cite{HirataJha,HirataJha2} and canonical ensemble \cite{JhaHirata_canonical}, every 
important conclusion for the former holds for the latter. 
Appendix \ref{app:derivation} documents a brief overview of the time-independent, 
algebraic derivations of the analytical 
formulas of $\Omega^{(n)}$, $U^{(n)}$, and $\mu^{(n)}$ ($0 \leq n \leq 2$), which serve as a basis of the analysis.  

\section{Zero-temperature limit of $U$\label{sec:U}} 

The SoS analytical formulas for the zeroth- \cite{Kou}, first- \cite{HirataJha}, and second-order \cite{HirataJha2}
perturbation corrections of $U$ are written as
\begin{eqnarray}
U^{(0)} &=& \langle E_I^{(0)} \rangle, \label{U0_SoS} \\
U^{(1)} 
&=& \langle E_I^{(1)} \rangle - \beta \langle F_I^{(0)}  F_I^{(1)}  \rangle \
+ \beta\langle F_I^{(0)} \rangle  \langle F_I^{(1)} \rangle  \label{U1_SoS}, \\
U^{(2)} &=& \langle E_I^{(2)} \rangle -{\beta} \langle F_I^{(1)} F_I^{(1)} \rangle
+{\beta} \langle F_I^{(1)} \rangle  \langle F_I^{(1)} \rangle 
\nonumber\\&& 
-{\beta} \langle  F_I^{(0)} F_I^{(2)} \rangle
+{\beta} \langle F_I^{(0)} \rangle \langle F_I^{(2)} \rangle 
\nonumber\\&& 
+\frac{\beta^2}{2} \langle F_I^{(0)} (F_I^{(1)})^2\rangle
-\frac{\beta^2}{2} \langle F_I^{(0)} \rangle \langle (F_I^{(1)})^2 \rangle  
\nonumber\\&& 
-{\beta^2} \langle F_I^{(0)} F_I^{(1)}  \rangle\langle F_I^{(1)} \rangle
+{\beta^2} \langle F_I^{(0)} \rangle \langle F_I^{(1)}\rangle^2, \label{U2_SoS}
\end{eqnarray}
where $\langle X_I \rangle$ stands for the zeroth-order thermal average,
\begin{eqnarray}
\langle X_I \rangle = \frac{\sum_I X_I e^{-\beta F^{(0)}_I}}{\sum_I e^{-\beta F^{(0)}_I}},  \label{X}
\end{eqnarray}
with
\begin{eqnarray}
F_I^{(n)} = E_I^{(n)}-\mu^{(n)}N_I. \label{F}
\end{eqnarray}
Here, $\mu^{(n)}$ is the $n$th-order correction to the chemical potential, discussed fully in Sec.\ \ref{sec:mu}.
See Appendix \ref{app:derivation} for derivation of Eqs.\ (\ref{U0_SoS})--(\ref{U2_SoS}). 

\subsection{Nondegenerate, correct reference}

Let us first establish analytically that the finite-temperature perturbation theory passes the first KL test [Eq.\ (\ref{KL_test})]
for a nondegenerate, correct reference. 
A ``nondegenerate'' reference means that the degree of degeneracy of the reference (which can be higher than one) stays the same up to the relevant perturbation order.
By ``correct,'' we demand that the reference wave function morphs into 
the true ground-state wave function of FCI as $\lambda = 0 \to 1$. These correspond to the case in Fig.\ \ref{fig:1}.

Under these conditions, we can identify 
one and only one nondegenerate neutral reference state whose $F_0^{(0)}$ is the most negative. Then, each zeroth-order
thermal average $\langle X_I \rangle$ reduces to $X_0$ at $T = 0$. Also, a thermal average of products $\langle X_I Y_I \rangle$ becomes 
the single product $X_0 Y_0$ at $T=0$. Therefore, we have
\begin{eqnarray}
\lim_{T \to 0} U^{(0)} &=& E_0^{(0)}, \label{U0_SoS_limit_nondeg} \\
\lim_{T \to 0} U^{(1)} 
&=&  E_0^{(1)}  - \beta  F_0^{(0)}  F_0^{(1)}   
 + \beta F_0^{(0)}   F_0^{(1)}   \nonumber \\
&=&  E_0^{(1)}, \label{U1_SoS_limit_nondeg} \\
\lim_{T \to 0} U^{(2)} 
&=&  E_0^{(2)}  -{\beta}  F_0^{(1)} F_0^{(1)} 
+{\beta}  F_0^{(1)}  F_0^{(1)} 
\nonumber\\&& 
-{\beta}  F_0^{(0)} F_0^{(2)} 
+{\beta}  F_0^{(0)}  F_0^{(2)} 
\nonumber\\&& 
+\frac{\beta^2}{2}  F_0^{(0)} (F_0^{(1)})^2
-\frac{\beta^2}{2}  F_0^{(0)}   (F_0^{(1)})^2   
\nonumber\\&& 
-{\beta^2}  F_0^{(0)} (F_0^{(1)})^2
+{\beta^2}  F_0^{(0)}   (F_0^{(1)})^2
\nonumber\\
&=& E_0^{(2)},  \label{U2_SoS_limit_nondeg}
\end{eqnarray}
satisfying Eq.\ (\ref{KL_test}) for $0 \leq n \leq 2$.
This conclusion was numerically verified also \cite{HirataJha,HirataJha2}.
We can say that the Kohn--Luttinger conundrum does not exist for a nondegenerate, correct reference.

The internal energy formulas in the canonical ensemble \cite{JhaHirata_canonical} are the same as Eqs.\ (\ref{U0_SoS})--(\ref{U2_SoS})
with each $F_I^{(n)}$ replaced by $E_I^{(n)}$, 
also in the definition of the thermal average $\langle X_I \rangle$ [Eq.\ (\ref{X})]. 
Hence, they also pass the first KL test [Eq.\ (\ref{KL_test})] for a nondegenerate, correct reference
for up to the third order \cite{JhaHirata_canonical}.


\subsection{Degenerate and/or incorrect reference}

\begin{figure*}
\includegraphics[width=0.95\textwidth]{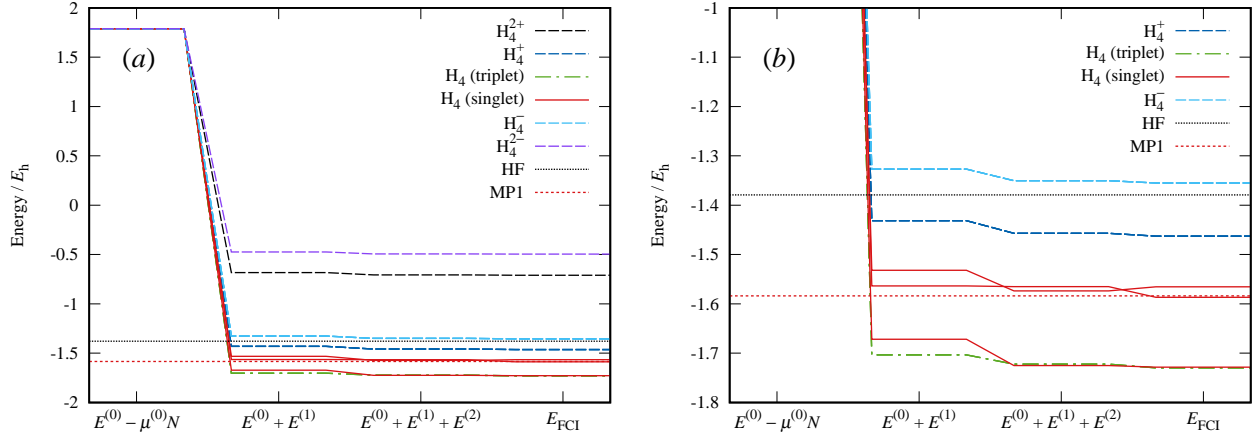}
\caption{({\it a}) The HCPT and FCI energies of the sixteen states sharing the same lowest $E_I^{(0)}-\mu^{(0)}N_I$
of the square-planar H$_4$ molecule (0.8~\AA) in the STO-3G basis set. HF stands for the zero-temperature limit
of the finite-temperature Hartree--Fock energy for the neutral, singlet ground state. MP1 refers 
to the first-order M{\o}ller--Plesset perturbation energy
for the neutral, singlet ground state. See footnotes of Table \ref{tab:deg_U} for more details.
({\it b}) A close up of ({\it a}). The neutral, triplet ground state (dotted-dashed green lines) is the overall ground state according to FCI.
\label{fig:HCPTenergies}}
\end{figure*}


The square-planar H$_4$ molecule \cite{H4} (with the side length of 0.8~\AA\ in the minimal basis set) is chosen
as a smallest system that has a degenerate and incorrect reference as it belongs to the non-Abelian point group of D$_{4h}$.
The reference is the zero-temperature limit of the finite-temperature canonical Hartree--Fock (HF) wave function 
for the neutral singlet ground state, whose
highest occupied molecular orbital (HOMO) and lowest unoccupied molecular orbital (LUMO) have the same energy.

Figure \ref{fig:HCPTenergies} plots the exact (FCI) energies of 
the sixteen zeroth-order degenerate states of the square-planar H$_4$ and their ions that have the same lowest $F_I^{(0)}$.
These figures also plot the zeroth-, first-, and second-order HCPT energies of the sixteen states. 
It can be seen that the degeneracy is already lifted at the first order of HCPT, revealing 
which state is the true ground state whose energy becomes the correct zero-temperature limit of $U$. 

Of particular interest among these sixteen states 
are the six states of the neutral H$_4$ sharing the identical $E_I^{(0)}$ and also the same $N_I=4$. 
(The rest are the states of the ions 
with the same $F_I^{(0)}$ but different $N_I$.) Three of them 
are singlet states plotted in solid red lines, while the other three are a triplet state
drawn in dotted-dashed green lines. This triplet state is the true ground state according to FCI, obeying Hund's rule, 
although the lowest singlet state (solid red lines)
is the reference wave function used 
in the finite-temperature 
perturbation calculations as well as the HCPT calculations generating these plots.

Hence, the square-planar H$_4$ calculation with the singlet ground-state reference is not only an example of
the case discussed in Fig.\ \ref{fig:2} (the zeroth-order degeneracy is lifted at the first order), but 
 also of the case in Fig.\ \ref{fig:3} (the reference does not correspond to the true ground state).  
 Keeping this in mind, we will analyze the general 
 $T \to 0$ behaviors of $U^{(0)}$, $U^{(1)}$, and $U^{(2)}$  in the following.

The SoS formula of $U^{(0)}$ [Eq.\ (\ref{U0_SoS})] can be rewritten as
\begin{eqnarray}
U^{(0)} &=& \langle F_I^{(0)} \rangle + \mu^{(0)} \langle N_I \rangle \nonumber\\
&=& \langle F_I^{(0)} \rangle + \mu^{(0)} \bar{N}, \label{U0_SoS2}
\end{eqnarray}
where $\bar{N}$ is the average number of electrons that ensures the electroneutrality of the molecule \cite{JhaHirata,HirataJha,HirataJha2}, and the second equality
follows from the fact that $\mu^{(0)}$ is determined by the condition $\langle N_I \rangle=\bar{N}$.
As $T \to 0$, the zeroth-order thermal average [Eq.\ (\ref{X})] is increasingly dominated by the states with the lowest $F_I^{(0)}$
and becomes its simple average over the zeroth-order degenerate reference states at $T=0$ (all of the sixteen states drawn in Fig.\  \ref{fig:HCPTenergies} in our H$_4$ example). 
Since all the degenerate reference states share the same $F_I^{(0)}$, we infer
\begin{eqnarray}
\lim_{T \to 0} U^{(0)} &=& F_0^{(0)} + \mu^{(0)} \bar{N} = E_0^{(0)},  \label{U0_SoS_limit_deg}
\end{eqnarray}
where $E_0^{(0)}$ is the zeroth-order energy of the neutral reference state.
Therefore, the SoS analytical formula of $U^{(0)}$ reaches the correct zero-temperature limit of 
$E_0^{(0)}$ for a degenerate, incorrect reference insofar as the reference (the singlet ground state in our H$_4$ example) belongs to the same zeroth-order degenerate subspace 
with the true ground state (the triplet ground state in H$_4$). 

\begin{figure}
\includegraphics[width=0.475\textwidth]{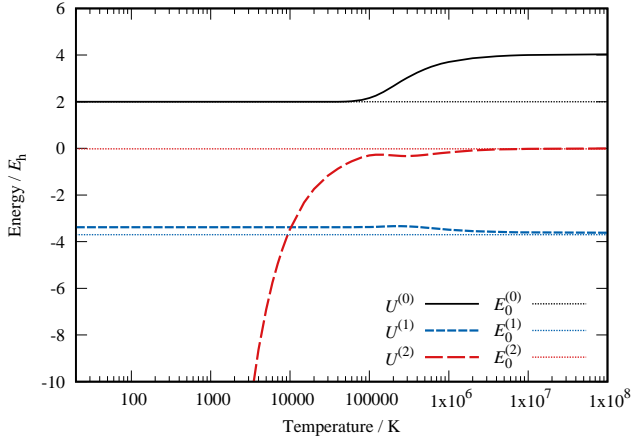}
\caption{The zeroth-, first-, and second-order perturbation corrections to the internal energy 
($U^{(0)}$, $U^{(1)}$, and $U^{(2)}$) as a function of temperature
as well as the HCPT energy corrections ($E_0^{(0)}$, $E_0^{(1)}$, and $E_0^{(2)}$) of the neutral triplet ground state 
as the correct zero-temperature limits. \label{fig:temperature}}
\end{figure}

This conclusion is verified numerically in Fig.\ \ref{fig:temperature}, whose selected data are compiled in Table \ref{tab:deg_U}. For the square-planar H$_4$ with the singlet reference which is degenerate with the triplet zeroth-order ground state, $U^{(0)}$ converges at its correct zero-temperature limit of $1.9980\,E_{\text{h}}$, which is equal to $E_0^{(0)}$ according to HCPT and MPPT as well as the corresponding energy component of the finite-temperature HF theory at $T=0$. 

\begin{table} 
\caption{\label{tab:deg_U} Comparison of the zeroth-, first-, and second-order corrections to  the internal energy ($U^{(n)}$, $0 \leq n \leq 2$)
as a function of temperature ($T$) for the square-planar H$_4$ molecule (0.8~\AA) in the STO-3G basis set.
The reference wave function is obtained as the zero-temperature limit of the finite-temperature Hartree--Fock calculation
for the singlet ground state and is degenerate.}
{
\begin{ruledtabular}
\begin{tabular}{lddd}
{$T /~\text{K}$} 
&\multicolumn{1}{c}{$U^{(0)} / E_{\text{h}}$} 
&\multicolumn{1}{c}{$U^{(1)} / E_{\text{h}}$} 
&\multicolumn{1}{c}{$U^{(2)} / E_{\text{h}}$} \\ \hline
0 (HCPT)\footnotemark[1] & 1.9980 & -3.7015 & -0.0187 \\
0 (HCPT)\footnotemark[2] & 1.9980 & -3.6696 & -0.0534 \\
0 (MPPT)\footnotemark[3] & 1.9980 & -3.5817 & \multicolumn{1}{c}{$-\infty$} \\
0 (HF)\footnotemark[4] & 1.9980 & -3.3771 & \multicolumn{1}{c}{$\cdots$} \\
$10^2$ & 1.9980 & -3.3771 & -343.9555 \\
$10^3$ & 1.9980 & -3.3771 & -34.4176 \\
$10^4$ & 1.9980 & -3.3771 & -3.4638 \\
$10^5$ & 2.1568 & -3.3690 & -0.3002 \\
$10^6$ & 3.7078 & -3.4831 & -0.1684 \\
\end{tabular}
\footnotetext[1]{The correct zero-temperature limit. The Hirschfelder--Certain degenerate perturbation theory \cite{Hirschfelder} for the triplet ground state. The FCI wave function of this state is
$-0.70(1a_{1g})^2(2e_u\alpha)^1(3e_u\beta)^1(4b_{1g})^0+0.70(1a_{1g})^2(2e_u\beta)^1(3e_u\alpha)^1(4b_{1g})^0$.}
\footnotetext[2]{The Hirschfelder--Certain degenerate perturbation theory \cite{Hirschfelder} for the singlet ground state. The FCI wave function of this state is
$-0.57(1a_{1g})^2(2e_u)^2(3e_u)^0(4b_{1g})^0+0.57(1a_{1g})^2(2e_u)^0(3e_u)^2(4b_{1g})^0
+0.40(1a_{1g})^2(2e_u\alpha)^1(3e_u\beta)^1(4b_{1g})^0+0.40(1a_{1g})^2(2e_u\beta)^1(3e_u\alpha)^1(4b_{1g})^0$.}
\footnotetext[3]{The M{\o}ller--Plesset perturbation theory \cite{moller} for the singlet
Slater-determinant reference: $(1a_{1g})^2(2e_u)^2(3e_u)^0(4b_{1g})^0$.}
\footnotetext[4]{The zero-temperature limit of the finite-temperature Hartree--Fock theory. 
The wave function is not a single Slater determinant, but is a linear combination of the form $2^{-1/2}(1a_{1g})^2(2e_u)^2(3e_u)^0(4b_{1g})^0+2^{-1/2}(1a_{1g})^2(2e_u)^0(3e_u)^2(4b_{1g})^0$.}
\end{ruledtabular}
}
\end{table}

The SoS analytical formula \cite{HirataJha,HirataJha2} of $U^{(1)}$  is given by Eq.\ (\ref{U1_SoS}).
The last two terms are alarming as they are individually divergent at $T = 0$; if $\langle F_I^{(0)}  F_I^{(1)}  \rangle$ and $\langle F_I^{(0)} \rangle  \langle F_I^{(1)} \rangle$ 
were not equal to each other at $T=0$, the zero-temperature limit of $U^{(1)}$ 
would be divergent and thus could not agree with the correct zero-temperature limit of $E_0^{(1)}$, which is always finite.

As $T \to 0$, each of these thermal averages is dominated by the simple average 
in the
zeroth-order degenerate subspace sharing the same lowest $F_I^{(0)}$. 
Generally, the degeneracy of $E_I^{(n)}$ is gradually lifted as the perturbation order $n$ is raised
and  hence the values of $E_I^{(1)}$ within the degenerate subspace usually have a distribution (as in our H$_4$ example as shown in Fig.\ \ref{fig:HCPTenergies}). 
Then, the sum of the last two terms becomes a covariance 
of two distributions, $F_I^{(0)}$ and $F_I^{(1)}$, multiplied by $-\beta$, i.e.,
\begin{eqnarray}
\beta \langle F_I^{(0)}  F_I^{(1)}  \rangle \
- \beta  \langle F_I^{(0)} \rangle  \langle F_I^{(1)} \rangle = \beta\, \text{cov}\left(F_I^{(0)},F_I^{(1)}\right),
\end{eqnarray}
at $T=0$.
In this case, however, $F_I^{(0)}$ has the same lowest value across all of
the degenerate
states and thus zero variance, and hence,
\begin{eqnarray}
\lim_{T \to 0} U^{(1)} &=& \lim_{T \to 0}  \langle E_I^{(1)} \rangle \equiv E\left[E_I^{(1)}\right], \label{U1_SoS_limit_deg}
\end{eqnarray}
where $E\left[X_I\right]$ stands for the simple average of $X_I$ over the zeroth-order degenerate states. This limit is finite. 

Does this mean that $U^{(1)}$ passes the first KL test [Eq.\ (\ref{KL_test})] for a degenerate reference? 
The answer is no because the simple average of $E_I^{(1)}$ within the zeroth-order degenerate subspace is different from $E_0^{(1)}$ for the true ground state, the latter being the correct zero-temperature limit. We, therefore, conclude
\begin{eqnarray}
\lim_{T \to 0} U^{(1)} = E\left[E_I^{(1)}\right] \neq E_0^{(1)}, \label{U1_SoS_limit_deg2}
\end{eqnarray}
indicating that although the first-order perturbation theory remains finite and well defined as $T \to 0$, it fails to converge at the correct zero-temperature limit when the degeneracy is partially or fully lifted at the first order of HCPT \footnote{In our previous studies \cite{HirataJha2,JhaHirata_canonical}, we argued that the physically correct way of taking the thermal average of $E_I^{(1)}$
at $T=0$ is to give 100\% weight to the lowest $E_I^{(1)}$, so that $U^{(1)}$ tends to the lowest $E_I^{(1)}$ as $T \to 0$, allowing the SoS formula for $U^{(1)}$ to pass the first KL test. 
This argument is troublesome for two reasons. First, at infinitesimal temperature ($T=0^+$), the weight is  constant across all degenerate states, making 
$U^{(1)}$ jump from the simple average of  $E_I^{(1)}$ in the degenerate subspace to its lowest value as $T = 0^+ \to 0$, which is both nonphysical (qualitatively different from experimental reality) and 
nonmathematical (not meeting the mathematical condition of 
a limit). Second, the lowest $E_I^{(1)}$ may not correspond to the true ground state of FCI, 
i.e., when the case of Fig.\ \ref{fig:3} applies, as in our H$_4$ example,
where the lowest $E_I^{(1)}$ is still not the correct zero-temperature limit, $E_0^{(1)}$.}.

According to Table \ref{tab:deg_U} \footnote{The two values of the first-order HCPT energy corrections are the two distinct eigenvalues of the perturbation matrix [Eq.\ (37) of Ref.\ \onlinecite{Hirschfelder}] within the degenerate subspace, corresponding 
to the triplet and singlet neutral ground states. The first-order MPPT energy correction is obtained by evaluating the well-known formula [Eq.\ (B1) of Ref.\ \onlinecite{HirataJha2}]
for the single Slater determinant for the singlet neutral ground state. In the zero-temperature limit, the finite-temperature HF theory imparts equal weights (via the density matrix) to the two Slater determinants for the singlet neutral ground state 
that are symmetrically and energetically equivalent. Its energy minus the zeroth-order energy [Eq.\ (\ref{U0_reduced_limit_deg})] is listed as 
the first-order correction according to the finite-temperature HF theory at $T=0$. The last method generated the degenerate  reference for the neutral singlet ground state.},
the zero-temperature limit of $U^{(1)}$ in the square-planar H$_4$ is $-3.3771\,E_{\text{h}}$ (reached at $T=10^2\,\text{K}$) and is
distinctly higher than the correct zero-temperature limit of $E_0^{(1)}=-3.7015\,E_{\text{h}}$, which is the first-order HCPT energy correction for the 
neutral triplet ground state, supporting the above conclusion [Eq.\ (\ref{U1_SoS_limit_deg2})] numerically. 
Figure \ref{fig:temperature} shows the same graphically.


The $U^{(1)}$ expression in the canonical ensemble \cite{JhaHirata_canonical} is the same as Eq.\ (\ref{U1_SoS}) with every $F_I^{(n)}$ replaced by $E_I^{(n)}$. 
Hence, the same conclusion holds:\ The first-order perturbation theory in the canonical ensemble fails the first KL test when 
the reference is degenerate and/or incorrect. 

The zero-temperature limit of the SoS analytical formula for $U^{(2)}$ [Eq.\ (\ref{U2_SoS})] is
\begin{eqnarray}
\lim_{T \to 0} U^{(2)} &=& E\left[  E_I^{(2)} \right] -{\beta}\,\text{cov} \left(F_I^{(1)}, F_I^{(1)}\right)
\nonumber\\&& 
-{\beta}\,\text{cov}\left( F_I^{(0)}, F_I^{(2)} \right)
+\frac{\beta^2}{2}\,\text{cov} \left( F_I^{(0)}, (F_I^{(1)})^2 \right)
\nonumber\\&& 
-{\beta^2}\,\text{cov}\left( F_I^{(0)}, F_I^{(1)} \right) E\left[ F_I^{(1)} \right] \\
&=& E\left[  E_I^{(2)} \right]  -{\beta}\,\text{cov} \left(F_I^{(1)}, F_I^{(1)}\right), \label{U2_SoS_limit_deg}
\end{eqnarray}
where the simple average  and covariance  are taken over all zeroth-order degenerate states. 
In general, $F_I^{(1)}$ has a lower degree of degeneracy than $F_I^{(0)}$, whence it has a nonzero variance, 
making Eq.\ (\ref{U2_SoS_limit_deg}) divergent as $T \to 0$. 
We can thus write
\begin{eqnarray}
\lim_{T \to 0} U^{(2)} = -\infty \neq E_0^{(2)}, \label{U2_SoS_limit_deg2}
\end{eqnarray}
when the degeneracy of the reference is partially or fully lifted at the first order of HCPT.

If the degree of degeneracy remains unchanged at the first order but is lowered at the second order, 
$U^{(2)}$ converges at a finite, but wrong zero-temperature limit because $E\left[  E_I^{(2)} \right]$ differs from $E_0^{(2)}$:
\begin{eqnarray}
\lim_{T \to 0} U^{(2)} = E\left[ E_I^{(2)} \right] \neq E_0^{(2)}. \label{U2_SoS_limit_deg3}
\end{eqnarray}
If the degree of degeneracy 
stays the same up to the second order, $U^{(2)}$ converges at  
the correct zero-temperature limit of $E_0^{(2)}$ provided the reference is correct.

Table \ref{tab:deg_U} and Fig.\ \ref{fig:temperature} numerically verify the above conclusion for the square-planar H$_4$. 
The correct zero-temperature limit of $U^{(2)}$ is the second-order HCPT energy correction for the neutral triplet ground state,
which is $-0.0187\,E_{\text{h}}$, whereas $U^{(2)}$ becomes asymptotically inversely proportional to $T$ and tends 
to $-\infty$ as $T \to 0$. The second-order MPPT energy correction in the square-planar H$_4$ 
is also $-\infty$, but this superficial agreement ($-\infty=-\infty$) merely constitutes a misuse of the nondegenerate MPPT  for a degenerate reference.

Since $U^{(2)}$ in the canonical ensemble \cite{JhaHirata_canonical} is isomorphic to Eq.\ (\ref{U2_SoS}), 
it also suffers from divergence at $T=0$ if the degeneracy is lifted at the first order. 
While the divergence in the grand canonical ensemble might possibly (though highly improbably) be systematically
removed by a clever choice of $\mu$ \cite{kohn,luttingerward,balian}, it does not fundamentally 
resolve the Kohn--Luttinger conundrum 
because the divergence persists in the canonical ensemble, which does not involve $\mu$. 


See Appendix \ref{app:U} for the analysis based on the reduced analytical formulas of $U^{(0)}$, $U^{(1)}$, and $U^{(2)}$,
leading to the same conclusions.

\section{Zero-temperature limit of $\mu$\label{sec:mu}}

\subsection{Nondegenerate, correct references\label{sec:mu_nondeg}}

Here, the ``nondegenerate, correct'' references pertain to all of the neutral, cation, and anion ground states.
The cation reference is the one in which an electron in HOMO is annihilated from 
the neutral reference. The anion reference  is the one in which an electron is created in LUMO of the neutral reference. By ``nondegenerate,'' we mean that the degrees of degeneracy of these cation and anion references remain the same up to the relevant perturbation order and that the LUMO energy ($\epsilon_l$) is higher than the HOMO 
energy  ($\epsilon_{h}$). The ``correct'' cation and anion references 
are the ones that morph into the true cation and anion ground-state wave functions as $\lambda = 0 \to 1$. 

The reduced (sum-over-orbitals) equation to be solved for $\mu^{(0)}$ is \cite{HirataJha,HirataJha2}
\begin{eqnarray}
\bar{N} = \sum_p f_p^-, \label{mu0_reduced}
\end{eqnarray}
where $f_p^- = 1/\{ 1 + e^{\beta(\epsilon_p - \mu^{(0)})}\}$ is the Fermi--Dirac distribution function \cite{HirataJha,HirataJha2} and $p$ runs over all spinorbitals.
This equation becomes indeterminate at $T=0$ since it is satisfied by any $\mu^{(0)}$ in the range $\epsilon_h < \mu^{(0)} < \epsilon_l$.
However, at $T \approx 0$, the equality is ensured largely by 
the contributions from HOMO and LUMO only, satisfying 
\begin{eqnarray}
N_h^\text{deg.} f_h^+ = N_l^\text{deg.} f_l^-,
\end{eqnarray}
where $f_p^+ = 1 - f_p^-$ and $N_h^\text{deg.}$ and $N_l^\text{deg.}$ are
the degrees of degeneracy of HOMO and LUMO, respectively.
This can be solved for $\mu^{(0)}$ to give
\begin{eqnarray}
\mu^{(0)} = \frac{\epsilon_{h} + \epsilon_{l}}{2} + \frac{1}{2\beta} \ln\frac{N^{\text{deg.}}_h}{N^{\text{deg.}}_l},
\end{eqnarray}
at $T \approx 0$, which implies \cite{Kou}
\begin{eqnarray}
\lim_{T \to 0} \mu^{(0)} = \frac{\epsilon_{h} + \epsilon_{l}}{2}. \label{mu0_0}
\end{eqnarray} 

On the other hand, the correct zero-temperature limit of $\mu^{(n)}$ [the right-hand side of Eq.\ (\ref{KL_test2})] for 
nondegenerate, correct references can be further simplified as
\begin{eqnarray}
\frac{E^{(n)}_\text{anion} - E^{(n)}_\text{cation}}{2} = \frac{\Sigma^{(n)}_h + \Sigma^{(n)}_l}{2}, \label{Dyson}
\end{eqnarray}
where $\Sigma_p^{(n)}$ is the $\Delta$MP$n$ energy \cite{deltamp}
for the $p$th spinorbital, which is, in turn, equal to the Dyson self-energy in the diagonal and
frequency-independent approximation \cite{Hirata2017} for $1 \leq n \leq 3$. 

Since $\Sigma_p^{(0)} = \epsilon_p$ \cite{szabo,Hirata2017}, the Fermi--Dirac theory
passes the second KL test [Eq.\ (\ref{KL_test2})]
for nondegenerate, correct references:
\begin{eqnarray}
\lim_{T \to 0} \mu^{(0)} = \frac{\epsilon_{h} + \epsilon_{l}}{2} = \frac{E^{(0)}_\text{anion} - E^{(0)}_\text{cation}}{2}. \label{mu0_0_new}
\end{eqnarray} 

The reduced analytical formula of $\mu^{(1)}$ (see Appendix \ref{app:derivation}) is given \cite{HirataJha,HirataJha2} by 
\begin{eqnarray}
\mu^{(1)} = \frac{\sum_p F_{pp} f_p^- f_p^+ }{\sum_p f_p^- f_p^+}, \label{mu1_reduced}
\end{eqnarray}
where $\bm{F}$ is the finite-temperature Fock matrix \cite{SANTRA,HirataJha2} minus the diagonal zero-temperature Fock matrix,
\begin{eqnarray}
F_{pq} = H_{pq}^{\text{core}} + \sum_r \langle pr || qr \rangle f_r^- -\delta_{pq} \epsilon_p, \label{Fock}
\end{eqnarray}
with $\bm{H}^{\text{core}}$ being the one-electron
part of the Fock matrix \cite{szabo},
when the M{\o}ller--Plesset partitioning \cite{moller} of the Hamiltonian is employed, 
and $\langle pq || rs \rangle$ is the anti-symmetrized two-electron integral \cite{shavitt}.
Since the finite-temperature canonical HF wave function at $T=0$ is used as the reference, $F_{pq} = 0$ at $T=0$, leading 
us to conclude 
\begin{eqnarray}
\lim_{T \to 0} \mu^{(1)} =  0.\label{mu1_0_2}
\end{eqnarray}
In the meantime, the right-hand side of Eq.\ (\ref{KL_test2}) is also zero because 
$\Sigma_p^{(0)} = 0$ according to Koopmans' theorem \cite{szabo,Hirata2017}. 
For nondegenerate, correct references, therefore, the first-order perturbation theory also passes 
the second KL test, i.e.,
\begin{eqnarray}
\lim_{T \to 0} \mu^{(1)} = 0 =  \frac{E_\text{anion}^{(1)} - E_\text{cation}^{(1)}}{2}. \label{mu1_0_new}
\end{eqnarray}

The reduced analytical formula of $\mu^{(2)}$ (see Appendix \ref{app:derivation}) is written as \cite{HirataJha2}
\begin{widetext}
\begin{eqnarray}
\mu^{(2)} \sum_p f_p^- f_p^+  
&=& \sum_{p,q}^{\text{denom.}\neq 0} \frac{| F_{pq} |^2 f_p^- f_q^+(f_p^+ - f_q^-)}{\epsilon_p-\epsilon_q}
+ \sum_{p,q,r}^{\text{denom.}\neq 0} \frac{ (F_{qp} \langle pr||qr\rangle + \langle qr||pr\rangle F_{pq} ) f_p^- f_q^+ f_r^- f_r^+ }{\epsilon_p-\epsilon_q}
\nonumber\\&& 
+ \frac{1}{4} \sum_{p,q,r,s}^{\text{denom.}\neq 0} \frac{| \langle pq || rs \rangle |^2 f_p^- f_q^- f_r^+ f_s^+(f_p^+ + f_q^+ - f_r^- - f_s^-) }{\epsilon_p+\epsilon_q-\epsilon_r-\epsilon_s} 
- \frac{\beta}{2} \sum_{p,q}^{\text{denom.}= 0} {| F_{pq} |^2 f_p^- f_q^+ (f_p^+ - f_q^-)}
\nonumber\\&& 
- \frac{\beta}{2} \sum_{p,q,r}^{\text{denom.}= 0} {(F_{qp} \langle pr||qr\rangle + \langle qr||pr\rangle F_{pq} ) f_p^- f_q^+ f_r^- f_r^+ }
- \frac{\beta}{8} \sum_{p,q,r,s}^{\text{denom.}= 0} {| \langle pq || rs \rangle |^2 f_p^- f_q^- f_r^+ f_s^+(f_p^+ + f_q^+ - f_r^- - f_s^-)} 
\nonumber\\&& 
+ \beta \mu^{(1)} \sum_p F_{pp}f_p^- f_p^+ (f_p^+ - f_p^-)
+ \beta \mu^{(1)} \sum_{p,q} \langle pq || pq \rangle f_p^- f_p^+ f_q^-  f_q^+
- \frac{\beta}{2} \left(\mu^{(1)}\right)^2 \sum_p f_p^- f_p^+ (f_p^+ - f_p^-) , \label{mu2_reduced} 
\end{eqnarray}
where ``denom.$\neq$0'' means that the sum is taken over $p$ and $q$ that satisfy $\epsilon_p - \epsilon_q \neq 0$ or
over $p$, $q$, $r$, and $s$ that satisfy $\epsilon_p + \epsilon_q - \epsilon_r - \epsilon_s \neq 0$
(and ``denom.=0'' vice versa). 
At $T = 0$, $F_{pq} = 0$ and $\mu^{(1)}= 0$. For a neutral nondegenerate reference, the summations 
with the ``denom.=0'' restriction never take place, leaving 
\begin{eqnarray}
\lim_{T \to 0} \mu^{(2)}  
&=& \frac{1}{\lim_{T \to 0} \sum_p f_p^- f_p^+}  \lim_{T \to 0} \frac{1}{4} \sum_{p,q,r,s} \frac{| \langle pq || rs \rangle |^2 f_p^- f_q^- f_r^+ f_s^+(f_p^+ + f_q^+ - f_r^- - f_s^-) }{\epsilon_p+\epsilon_q-\epsilon_r-\epsilon_s} \\
&=& \frac{1}{4} \sum_{j,a,b} \frac{|\langle h j||ab\rangle|^2}{\epsilon_h  + \epsilon_j - \epsilon_a - \epsilon_b}
- \frac{1}{4} \sum_{i,j,a} \frac{|\langle ij||h a\rangle|^2}{\epsilon_i + \epsilon_j- \epsilon_h  - \epsilon_a} 
+ \frac{1}{4} \sum_{j,a,b} \frac{|\langle l j||ab\rangle|^2}{\epsilon_l + \epsilon_j - \epsilon_a - \epsilon_b}
- \frac{1}{4} \sum_{i,j,a} \frac{|\langle ij||la\rangle|^2}{\epsilon_i + \epsilon_j- \epsilon_l - \epsilon_a},\label{mu2_reduced2} 
\end{eqnarray}
\end{widetext}
where $i$ and $j$ run over spinorbitals occupied in the reference Slater determinant
and $a$ and $b$ over spinorbitals unoccupied, while $h$ and $l$ stand for HOMO and LUMO, respectively. 
In the second equality, we used the fact that the $p=h$ and $p=l$ summands decay 
most slowly and thus dominate $\sum_p f_p^- f_p^+$ as $T \to 0$.
The right-hand side of Eq.\ (\ref{mu2_reduced2}) is identified as the average of the $\Delta$MP2 energies \cite{deltamp,Hirata2017} for HOMO and LUMO because
\begin{eqnarray}
\Sigma_p^{(2)} &=& \frac{1}{2} \sum_{j,a,b} \frac{|\langle p j||ab\rangle|^2}{\epsilon_p  + \epsilon_j - \epsilon_a - \epsilon_b}
+ \frac{1}{2} \sum_{i,j,a} \frac{|\langle ij||p a\rangle|^2}{\epsilon_p + \epsilon_a- \epsilon_i  - \epsilon_j}, \label{DeltaMP2}
\nonumber\\
\end{eqnarray}
proving
\begin{eqnarray}
\lim_{T \to 0} \mu^{(2)} =  \frac{\Sigma_{h}^{(2)} + \Sigma_{l}^{(2)}}{2} = \frac{E^{(2)}_\text{anion} - E^{(2)}_\text{cation}}{2}. \label{mu2_0_new}
\end{eqnarray}
Therefore, the second-order perturbation theory again passes the second KL test [Eq.\ (\ref{KL_test2})] for nondegenerate, correct references. 

\begin{table} 
\caption{\label{tab:deg_mu} Comparison of the zeroth-, first-, and second-order corrections to the chemical potential ($\mu^{(n)}$, $0 \leq n \leq 2$)
as a function of temperature ($T$) for the square-planar H$_4$ molecule (0.8~\AA) in the STO-3G basis set.
The HOMO and LUMO energies are $0.05235\,E_{\text{h}}$.}
{
\begin{ruledtabular}
\begin{tabular}{lddd}
{$T /~\text{K}$} 
&\multicolumn{1}{c}{$\mu^{(0)} / E_{\text{h}}$} 
&\multicolumn{1}{c}{$\mu^{(1)} / E_{\text{h}}$} 
&\multicolumn{1}{c}{$\mu^{(2)} / E_{\text{h}}$} \\ \hline
0\footnotemark[1] & 0.05235 & 0.00000 & 0.00086 \\
$10^2$ & 0.05235 & 0.00000 & 0.00086 \\
$10^3$ & 0.05235 & 0.00000 & 0.00086 \\
$10^4$ & 0.05235 & 0.00000 & 0.00086 \\
$10^5$ & 0.06832 & -0.00227 & 0.02292 \\
$10^6$ & 0.11259 & 0.00740 & 0.00013 \\
\end{tabular}
\footnotetext[1]{Equations (\ref{mu0_0_new}), (\ref{mu1_0_new}), and (\ref{mu2_0_new}). In the latter,
the summands with a vanishing denominator in Eq.\ (\ref{DeltaMP2}) were excluded.}
\end{ruledtabular}
}
\end{table}

The square-planar H$_4$ molecule with the neutral singlet reference generates the nondegenerate, correct references 
for the cation and anion. The cation reference is four-fold degenerate at any perturbation order and converges at the true cation ground state (see Fig.\ \ref{fig:HCPTenergies}). The same applies to the anion. However, the neutral singlet reference is degenerate (and the degeneracy is lifted at the first order) and is also incorrect (the true ground state is triplet). Therefore, strictly speaking,
 H$_4$ does not satisfy all of the conditions of nondegenerate, correct references.
Nevertheless, as Table \ref{tab:deg_mu} indicates, $\mu^{(0)}$, $\mu^{(1)}$, and $\mu^{(2)}$ all 
come within $0.1\,\text{m}E_\text{h}$ of the 
correct zero-temperature limits [Eqs.\ (\ref{mu0_0_new}), (\ref{mu1_0_new}), and (\ref{mu2_0_new})]
at $T \leq 10^4\,\text{K}$. 
This means that, under certain circumstances, 
the energy difference, $E^{(n)}_\text{anion} - E^{(n)}_\text{cation}$, can still be computed correctly
with a degenerate and/or incorrect neutral reference since the latter does not explicitly enter the difference formula.
In this case, however, Eq.\ (\ref{DeltaMP2}) 
needed to be adjusted so as to exclude the summands with a vanishing denominator, which is, in turn,
justified by a sum rule for the second-order HCPT energy corrections [cf.\ Eq.\ (B4) of Ref.\ \onlinecite{HirataJha2}].

\subsection{Degenerate and/or incorrect references}

If the degree of degeneracy of the cation or anion ground state is partially or fully lifted, $E^{(n)}_\text{cation}$ or $E^{(n)}_\text{anion}$ ($n \geq 1$) is only procedurally defined by HCPT as an eigenvalue of some perturbation matrix 
[e.g., Eqs.\ (37) and (57) of Ref.\ \onlinecite{Hirschfelder}] and cannot be 
written in a closed analytical formula or diagrammatically; Eq.\ (\ref{Dyson}) no longer holds. 
Furthermore, if the neutral ground state is degenerate, the $\Delta$MP$n$
expressions become ill-posed, making, e.g., Eq.\ (\ref{DeltaMP2}) divergent. 
If the cation or anion reference does not correspond to the respective true ground state, clearly 
the reduced formula (and its equivalent SoS formula) of $\mu^{(n)}$ converges
at a wrong zero-temperature limit. 
In short, the first- and higher-order perturbation theories generally fail the second KL test [Eq.\ (\ref{KL_test2})] 
for the cases that do not satisfy the conditions stipulated in the beginning of Sec.\ \ref{sec:mu_nondeg}. 

The Fermi--Dirac theory, on the other hand, passes the second KL test barring the most pathological cases. One such case
is when the energy ordering of the cation or anion ground state changes as $\lambda = 0 \to 1$.

\section{Conclusions}

Our findings are summarized as follows:

(1) The first-order perturbation corrections to the internal energy ($U$) and grand potential ($\Omega$)
according to the finite-temperature perturbation theory in the grand canonical ensemble approach wrong limits 
as $T \to 0$ and, therefore, become increasingly inaccurate at low temperatures when the reference is degenerate and/or
incorrect. The reference is considered degenerate if the degree of degeneracy changes with the perturbation order
up to the corresponding order. The reference is incorrect if it does not smoothly connect to the true ground-state wave function as the perturbation strength ($\lambda$) is raised from zero to unity. In principle, one cannot know if 
the reference is correct until a FCI calculation is performed for all states. 

(2) The first-order perturbation corrections to $U$ and $\Omega$ in the grand canonical ensemble reach finite zero-temperature limits,
which are nonetheless wrong when the degeneracy of the reference is lifted at the first order of HCPT or the reference is incorrect.

(3) The second-order perturbation corrections to $U$ and $\Omega$ in the grand canonical ensemble are 
divergent when the degeneracy of the reference is lifted at the first order. Otherwise
they converge at finite, but still wrong limits if the degeneracy is lifted at the second order or the reference is incorrect.

(4) The zeroth-order Fermi--Dirac theory in the grand canonical ensemble is much more robust and is 
correct in most (but not all) cases. 

(5) The zeroth-, first-, and second-order perturbation corrections to the chemical potential ($\mu$)
converge at the correct zero-temperature limits if all of the neutral, cation, and anion references are correct and their degrees of degeneracy
remain unchanged up to the corresponding perturbation order.
(The condition for the neutral reference may be  relaxed.)

(6) Conclusions (1) through (5) have been numerically verified for the square-planar H$_4$, which has a degenerate and 
incorrect neutral reference wave function.

(7) The zeroth-, first-, and second-order perturbation corrections to the internal energy and Helmholtz energy according
to the finite-temperature perturbation theory in the canonical ensemble display the same $T\to 0$ behaviors
as their counterparts in the grand canonical ensemble. 

(8) Taken together, the finite-temperature perturbation theory in the grand canonical and canonical ensembles 
has zero radius of convergence at $T = 0$ and becomes increasingly useless or even misleading at low temperatures when
the reference is degenerate and/or incorrect. Since this occurs in the canonical ensemble also, this problem 
cannot be resolved by a clever choice of $\mu$ contrary to some earlier propositions \cite{kohn,luttingerward,balian}. Rather, it originates from the nonanalyticity of the Boltzmann factor 
at $T=0$, preventing the energy expression from being expanded in a converging power series. Worse still, one cannot
know without carrying out a FCI calculation whether the degree of degeneracy remains the same up to FCI and whether the reference corresponds to the true ground state. Therefore, this conundrum exposes 
a particularly severe flaw of perturbation theory. 

\acknowledgments

This work was supported by the Center for Scalable, Predictive methods for Excitation and Correlated phenomena (SPEC), which is funded by 
the U.S. Department of Energy, Office of Science, Office of Basic Energy Sciences, 
Chemical Sciences, Geosciences, and Biosciences Division, as a part of the Computational Chemical Sciences Program
and also by the U.S. Department of Energy, Office of Science, Office of Basic Energy Sciences under Grant No.\ DE-SC0006028.

\appendix

\section{Justification of Eq.\ (\ref{KL_test2})\label{app:mu_limit}}
The chemical potential $\mu$ is determined by solving the electroneutrality condition \cite{JhaHirata,HirataJha,HirataJha2},
\begin{eqnarray}
\bar{N} =  \frac{\sum_I N_I e^{-\beta F_I}}{\sum_I e^{-\beta F_I}}, \label{neutrality}
\end{eqnarray}
where $\bar{N}$ is the average number of electrons that keeps the system electrically neutral. 
As $T \to 0$, the thermal average is increasingly dominated by the term with the most negative $F_I$,
where the $I$th state is usually the neutral (degenerate or nondegenerate) ground state (i.e., $I = 0$). However, if we
kept only this greatest summand in the numerator, we could not determine $\mu$ because
the equation would hold for any value of $\mu$. What actually determines
$\mu$ at $T \approx 0$ is the most dominant summands for ionized and electron-attached states with $N_I \neq \bar{N}$.
Assuming the most common scenario in which the most negative $F_I$ for ionized and electron-attached states 
occur for $N_I = \bar{N} \pm 1$, we see that the above equation is satisfied at $T=0$ if 
the contributions to the right-hand side from the cation and anion ground states 
cancel with each other exactly, i.e.,
\begin{eqnarray}
N^{\text{deg.}}_\text{cation} e^{-\beta E_{\text{cation}} +\beta \mu (\bar{N} - 1)} = N^{\text{deg.}}_\text{anion} e^{-\beta E_{\text{anion}}+\beta \mu (\bar{N} + 1)} ,
\end{eqnarray} 
where $E_\text{cation}$ and $N^{\text{deg.}}_\text{cation}$ are the energy and
degeneracy of the cation ground state (and the anion counterparts similarly defined). 
This can be solved for $\mu$ as 
\begin{eqnarray}
\mu &=& \frac{E_\text{anion} - E_\text{cation}}{2} + \frac{1}{2\beta} \ln\frac{N^{\text{deg.}}_\text{cation}}{N^{\text{deg.}}_\text{anion}},
\end{eqnarray}
at $T \approx 0$, which implies
\begin{eqnarray}
\lim_{T \to 0} \mu &=& \frac{E_\text{anion} - E_\text{cation}}{2}. \label{mu_limit}
\end{eqnarray}
Differentiating this equation with respect to $\lambda$, we recover Eq.\ (\ref{KL_test2}).

\section{The $T \to 0$ behavior of the reduced analytical formulas of $U^{(n)}$\label{app:U}}

The SoS (sum-over-states) and reduced (sum-over-orbitals) analytical formulas are mathematically equivalent
to each other, and hence the analysis based on the latter, given in this section, 
would merely confirm the conclusions drawn in the main body of this article, but it shines some light 
on the anomalous diagrams \cite{kohn}.

The reduced analytical formula for $U^{(0)}$ reads \cite{HirataJha,HirataJha2}
\begin{eqnarray}
U^{(0)} = E_{\text{nuc.}} + \sum_p \epsilon_p f_p^-, \label{U0_reduced}
\end{eqnarray}
where $E_{\text{nuc.}}$ is the nuclear-repulsion energy and $\epsilon_p$ is the canonical HF energy of 
the $p$th spinorbital, and the summation is taken over
all spinorbitals. At $T=0$, $f_p^-=1$ for all $p$ with $\epsilon_p \leq \epsilon_h$, and $f_p^- = 0$ for all $p$ with $\epsilon_p > \epsilon_h$, as well as (see also Ref.\ \onlinecite{PedersonJackson})
\begin{eqnarray}
\lim_{T \to 0}f_h^-=\lim_{T \to 0}f_l^- = \frac{N_h^\text{deg.}}{N_h^\text{deg.} + N_l^\text{deg.}},
\label{f_limits}
\end{eqnarray}
where $h$ stands for HOMO
and $l$ for LUMO, and $N_h^\text{deg.}$ 
and $N_l^\text{deg.}$ are the degrees of degeneracy of these spinorbitals.
Substituting, we obtain
\begin{eqnarray}
\lim_{T \to 0} U^{(0)} = E_{\text{nuc.}} + \sum_i^{\text{occ.}} \epsilon_i , \label{U0_reduced_limit_deg}
\end{eqnarray}
where `occ.'\ means that $i$ runs over spinorbitals occupied in the reference. 
The right-hand side is identified as the reduced analytical formula of $E_0^{(0)}$ \cite{HirataJha,HirataJha2}. Therefore, the Fermi--Dirac theory 
passes the first KL test [Eq.\ (\ref{KL_test})] in all cases except when the energy ordering of the ground state changes with $\lambda$. 

The reduced analytical formula of $U^{(1)}$ (see Appendix \ref{app:derivation}) reads \cite{HirataJha,HirataJha2}
\begin{eqnarray}
U^{(1)} 
&=& \sum_p F_{pp} f_p^- -\frac{1}{2} \sum_{p,q} \langle pq || pq \rangle f_p^- f_q^- 
\nonumber \\&& 
- \beta \sum_p  F_{pp} \epsilon_p   f_p^- f_p^+
+ \beta  \mu^{(1)} \sum_p \epsilon_p   f_p^- f_p^+, \label{U1_reduced}
\end{eqnarray}
where $\mu^{(1)}$ is given by Eq.\ (\ref{mu1_reduced}).
Taking the zero-temperature limit, we obtain
\begin{eqnarray}
\lim_{T \to 0} U^{(1)} 
&=& E\left[ \sum_i^{\text{occ.}} F_{ii}  \right]
- E\left[  \frac{1}{2}  \sum_{i,j}^{\text{occ.}} \langle ij || ij \rangle \right]
\nonumber \\&& 
- \beta \sum_p^{\epsilon_p=\epsilon_h}  F_{pp} \epsilon_p   f_p^- f_p^+
\nonumber \\&& 
+ \beta  \frac{\sum_p^{\epsilon_p=\epsilon_h} F_{pp} f_p^- f_p^+ }{\sum_p^{\epsilon_p=\epsilon_h} f_p^- f_p^+} 
\sum_p^{\epsilon_p=\epsilon_h} \epsilon_p   f_p^- f_p^+ \label{U1_reduced_limit_deg2} \\
&=& -E\left[ \frac{1}{2} \sum_{i,j}^{\text{occ.}} \langle ij || ij \rangle \right], \label{U1_reduced_limit_deg}
\end{eqnarray}
where $\epsilon_p = \epsilon_h$ means that $p$ runs over all spinorbitals that are degenerate with HOMO. The second equality used the fact that 
at $T = 0$, $f_p^- f_p^+=0$ for all $p$ but degenerate HOMO and LUMO whose $f_p^- f_p^+$ share some nonzero value [Eq.\ (\ref{f_limits})] as well as $\lim_{T\to 0} F_{pp} = 0$ as per Eq.\ (\ref{Fock}). 

For a nondegenerate, correct reference, Eq.\ (\ref{U1_reduced_limit_deg}) is an average of just one term and equals to 
the first-order MPPT energy correction \cite{szabo,shavitt} for the reference, which is the correct zero-temperature limit; the first-order perturbation theory passes the first KL test [Eq.\ (\ref{KL_test})]. When the degeneracy of the reference is lifted at the first order, the average of the first-order HCPT energy corrections within the degenerate subspace is no longer the same as
the first-order HCPT energy correction for the true ground state; the first-order perturbation theory fails the test.
When the reference is incorrect, the average has nothing to do with the correct zero-temperature limit and the theory again fails the test.

The penultimate term of Eq.\ (\ref{U1_reduced_limit_deg2}) contains 
\begin{eqnarray}
- \beta \sum_p^{\epsilon_p=\epsilon_h} \sum_{r} \langle pr || pr \rangle f_r^- \epsilon_p   f_p^- f_p^+,
\end{eqnarray}
which is divergent as $T \to 0$ and may be viewed as an anomalous contribution of Kohn and Luttinger \cite{kohn}
(although the parent term vanishes because $F_{pp} = 0$ at $T=0$).
That this is exactly canceled by the corresponding contribution in the last term containing $\mu^{(1)}$ appears 
to support the Luttinger--Ward prescription \cite{kohn,luttingerward,balian} even for a general, nonisotropic system. However, this cancellation only saves $U^{(1)}$
from divergence, and Eq.\ (\ref{U1_reduced_limit_deg}) still fails the first KL test [Eq.\ (\ref{KL_test})]
as already established above.
Therefore, whereas the first-order finite-temperature perturbation theory is not divergent thanks to this cancellation, it 
still tends to a wrong zero-temperature limit. The Luttinger--Ward prescription has a rather limited scope.

The reduced analytical formula of $U^{(2)}$ (see Appendix \ref{app:derivation}) reads \cite{HirataJha2}
\begin{widetext}
\begin{eqnarray}
U^{(2)} 
&=& \sum_{p,q}^{\text{denom.}\neq 0} \frac{| F_{pq} |^2 f_p^- f_q^+}{\epsilon_p-\epsilon_q} 
+ \frac{1}{4} \sum_{p,q,r,s}^{\text{denom.}\neq 0} \frac{| \langle pq || rs \rangle |^2 f_p^- f_q^- f_r^+ f_s^+}{\epsilon_p+\epsilon_q-\epsilon_r-\epsilon_s} 
- {\beta} \sum_{p,q}^{\text{denom.}= 0} {| F_{pq} |^2 f_p^- f_q^+}
- \frac{\beta}{4} \sum_{p,q,r,s}^{\text{denom.}= 0} {| \langle pq || rs \rangle |^2 f_p^- f_q^- f_r^+ f_s^+} 
\nonumber\\&& 
+ {\beta} \left(\mu^{(1)}\right)^2 \sum_p f_p^- f_p^+
- \beta \sum_{p,q}^{\text{denom.}\neq 0} \frac{| F_{pq} |^2 f_p^- f_q^+ (\epsilon_pf_p^+ - \epsilon_qf_q^-)}{\epsilon_p-\epsilon_q}
- \beta \sum_{p,q,r}^{\text{denom.}\neq 0} \frac{ (F_{qp} \langle pr||qr\rangle + \langle qr||pr\rangle F_{pq} )  f_p^- f_q^+  (\epsilon_r f_r^- f_r^+) }{\epsilon_p-\epsilon_q}
\nonumber\\&& 
- \frac{\beta}{4} \sum_{p,q,r,s}^{\text{denom.}\neq 0} \frac{| \langle pq || rs \rangle |^2 f_p^- f_q^- f_r^+ f_s^+
(\epsilon_pf_p^+ + \epsilon_qf_q^+ - \epsilon_rf_r^- - \epsilon_sf_s^-) }{\epsilon_p+\epsilon_q-\epsilon_r-\epsilon_s} 
+ \frac{\beta^2}{2} \sum_{p,q}^{\text{denom.}= 0} {| F_{pq} |^2 f_p^- f_q^+ (\epsilon_p f_p^+ - \epsilon_q f_q^-)}
\nonumber\\&& 
+ \frac{\beta^2}{2} \sum_{p,q,r}^{\text{denom.}= 0} {(F_{qp} \langle pr||qr\rangle + \langle qr||pr\rangle F_{pq} )  f_p^- f_q^+ (\epsilon_r f_r^-  f_r^+) }
+ \frac{\beta^2}{8} \sum_{p,q,r,s}^{\text{denom.}= 0} {| \langle pq || rs \rangle |^2 f_p^- f_q^- f_r^+ f_s^+(\epsilon_p f_p^+ + \epsilon_q f_q^+ - \epsilon_r f_r^- - \epsilon_s f_s^-)} 
\nonumber\\&& 
- \beta^2 \mu^{(1)} \sum_p F_{pp}f_p^- f_p^+ (\epsilon_p f_p^+ - \epsilon_p f_p^-)
- \beta^2 \mu^{(1)} \sum_{p,q} \langle pq || pq \rangle   f_p^- f_p^+ (\epsilon_q f_q^-  f_q^+)
\nonumber\\&& 
+ \frac{\beta^2}{2} \left(\mu^{(1)}\right)^2 \sum_p  f_p^-f_p^+ (\epsilon_p f_p^+ - \epsilon_pf_p^-) 
+ \beta \mu^{(2)} \sum_p \epsilon_p f_p^-f_p^+ ,
\label{U2_reduced} 
\end{eqnarray}
with $\mu^{(2)}$ given by Eq.\ (\ref{mu2_reduced}). In the zero-temperature limit, the last term with $\mu^{(2)}$ cancels
a majority of the remaining terms (the sixth through penultimate terms to be specific), leaving 
\begin{eqnarray}
\lim_{T \to 0} U^{(2)} 
&=& 
E\left[  \sum_{i,a}^{\text{denom.}\neq 0} \frac{| F_{ia} |^2}{\epsilon_i-\epsilon_a} \right]
+ E\left[  \frac{1}{4} \sum_{i,j,a,b}^{\text{denom.}\neq 0} \frac{| \langle ij || ab \rangle |^2 }{\epsilon_i+\epsilon_j-\epsilon_a-\epsilon_b} \right]
- {\beta} \sum_{p,q}^{\epsilon_p = \epsilon_q = \epsilon_h} {| F_{pq} |^2 f_p^- f_q^+}
- \frac{\beta}{4} \sum_{p,q,r,s}^{\epsilon_p = \epsilon_q =\epsilon_r = \epsilon_s = \epsilon_h} {| \langle pq || rs \rangle |^2 f_p^- f_q^- f_r^+ f_s^+} 
\nonumber\\&& 
+ {\beta} \left(\lim_{T \to 0} \mu^{(1)}\right)^2 \sum_p^{\epsilon_p = \epsilon_h} f_p^- f_p^+,
\label{U2_reduced_limit_deg} 
\end{eqnarray}
\end{widetext}
where the superscript ``$\text{denom.}\neq 0$'' 
excludes the summands with a vanishing denominator, while 
$\epsilon_p = \epsilon_h$, etc.\ mean that $p$ runs over all spinorbitals that are degenerate with HOMO.

For a nondegenerate, correct reference, each of the first two terms averages only one term and their sum is 
identified as the second-order MPPT energy 
correction \cite{szabo,shavitt} for the reference, which is the correct zero-temperature limit. The remaining three terms vanish, and, therefore, 
the second-order perturbation theory passes the first KL test [Eq.\ (\ref{KL_test})]. 

For a degenerate reference, the last three terms multiplied by $\beta$ generally do not cancel with one another at $T=0$,
causing $U^{(2)}$ to diverge. Even if it were not for these terms, the sum of the first two terms does not 
agree with the second-order HCPT energy correction for the true ground state, which is an eigenvalue of some 
perturbation matrix [Eq.\ (57) of Ref.\ \onlinecite{Hirschfelder}] and cannot be written
in a closed formula such as the above. 
Therefore, the second-order perturbation theory fails the first KL test for a degenerate reference.
It goes without saying that it fails when the reference is incorrect. 

\section{The $T \to 0$ behavior of the reduced analytical formulas of $\Omega^{(n)}$\label{app:Omega}}

The reduced analytical formula for $\Omega^{(0)}$ is given as \cite{Kou,HirataJha,HirataJha2}
\begin{eqnarray}
\Omega^{(0)} &=& E_\text{nuc.} + \frac{1}{\beta} \sum_p \ln f_p^+. \label{Omega0_reduced} 
\end{eqnarray}
For a nondegenerate, correct reference, we find
\begin{eqnarray}
\lim_{T \to 0} \Omega^{(0)} &=& E_\text{nuc.} +  \sum_i^{\text{occ.}} \left( \epsilon_i - \mu^{(0)} \right)
= E_0^{(0)} - \frac{\epsilon_h + \epsilon_l}{2} \bar{N}, \label{Omega0_limit_nondeg} 
\end{eqnarray}
where $i$ runs over all spinorbitals occupied in the reference, passing the third KL test [Eq.\ (\ref{KL_test3})].
For a degenerate, correct reference, using Eq.\ (\ref{f_limits}), we obtain
\begin{eqnarray}
\lim_{T \to 0} \Omega^{(0)} &=& E_\text{nuc.} +  \sum_i^{\epsilon_i < \epsilon_h} \left( \epsilon_i - \epsilon_h \right)
= E_0^{(0)} - {\epsilon_h} \bar{N}, \label{Omega0_limit_deg} 
\end{eqnarray}
again passing the third KL test because $\epsilon_h = \epsilon_l$. 

The reduced formula of $\Omega^{(1)}$  (see Appendix \ref{app:derivation}) reads \cite{HirataJha,HirataJha2}
\begin{eqnarray}
\Omega^{(1)} &=& \sum_p F_{pp} f_p^- -\frac{1}{2} \sum_{p,q} \langle pq || pq \rangle f_p^- f_q^- - \mu^{(1)} \bar{N}, \label{Omega1_reduced} 
\end{eqnarray}
where $\mu^{(1)}$ is given by Eq.\ (\ref{mu1_reduced}). To disentangle the $T \to 0$ behaviors of $\Omega$ and $\mu$, 
we henceforth assume that $\mu^{(n)}$ converges at the correct zero-temperature limit, which is denoted by $\lim_{T \to 0} \mu^{(n)}$. 
Using  $F_{pp} = 0$ at $T=0$, we obtain
\begin{eqnarray}
\lim_{T \to 0} \Omega^{(1)} &=&  -E \left[ \frac{1}{2} \sum_{i,j}^{\text{occ.}} \langle ij || ij \rangle \right] - \lim_{T \to 0} \mu^{(1)}\bar{N}. \label{Omega1_reduced_limit_deg} 
\end{eqnarray}
For a nondegenerate, correct reference, the first term is an average of just one term, which is 
identified as the first-order MPPT energy correction for the reference \cite{szabo,shavitt} 
and is the correct zero-temperature limit; the first-order perturbation theory passes the third KL test 
in this case. When the degeneracy  of the reference is lifted at the first order, the average differs from the 
first-order HCPT energy correction for the true ground state, and the theory fails the third KL test. For an incorrect reference,
the theory again fails to converge at the correct limit.

The reduced formula of $\Omega^{(2)}$  (see Appendix \ref{app:derivation}) reads \cite{HirataJha2}
\begin{eqnarray}
\Omega^{(2)} &=& \sum_{p,q}^{\text{denom.}\neq 0} \frac{| F_{pq} |^2 f_p^- f_q^+}{\epsilon_p-\epsilon_q}
+ \frac{1}{4} \sum_{p,q,r,s}^{\text{denom.}\neq 0} \frac{| \langle pq || rs \rangle |^2 f_p^- f_q^- f_r^+ f_s^+}{\epsilon_p+\epsilon_q-\epsilon_r-\epsilon_s} 
\nonumber\\&& 
- \frac{\beta}{2} \sum_{p,q}^{\text{denom.}= 0} {| F_{pq} |^2 f_p^- f_q^+}
\nonumber\\&& 
- \frac{\beta}{8} \sum_{p,q,r,s}^{\text{denom.}= 0} {| \langle pq || rs \rangle |^2 f_p^- f_q^- f_r^+ f_s^+} 
\nonumber\\&& 
+ \frac{\beta}{2} \left(\mu^{(1)}\right)^2 \sum_p f_p^- f_p^+ - \mu^{(2)} \bar{N} , \label{Omega2_reduced} 
\end{eqnarray}
where $\mu^{(2)}$ is given by Eq.\ (\ref{mu2_reduced}). Taking the zero-temperature limit, we find
\begin{eqnarray}
\lim_{T \to 0} \Omega^{(2)} 
&=& E\left[  \sum_{i,a}^{\text{denom.}\neq 0} \frac{| F_{ia} |^2}{\epsilon_i-\epsilon_a} \right]
\nonumber\\&& 
+   E\left[ \frac{1}{4} \sum_{i,j,a,b}^{\text{denom.}\neq 0} \frac{| \langle ij || ab \rangle |^2 }{\epsilon_i+\epsilon_j-\epsilon_a-\epsilon_b} \right]
\nonumber\\&&
- \frac{\beta}{2} \sum_{p,q}^{\epsilon_p = \epsilon_q = \epsilon_h} {| F_{pq} |^2 f_p^- f_q^+}
\nonumber\\&& 
- \frac{\beta}{8} \sum_{p,q,r,s}^{\epsilon_p = \epsilon_q =\epsilon_r = \epsilon_s = \epsilon_h} {| \langle pq || rs \rangle |^2 f_p^- f_q^- f_r^+ f_s^+} 
\nonumber\\&& 
+ \frac{\beta}{2} \left(\lim_{T \to 0} \mu^{(1)}\right)^2 \sum_p^{\epsilon_p = \epsilon_h} f_p^- f_p^+
-  \lim_{T \to 0} \mu^{(2)} \bar{N}.
\label{Omega2_reduced_limit_deg} 
\end{eqnarray}
The same mechanics are at play here as the $T\to 0$ behavior of $U^{(2)}$ (Appendix \ref{app:U}): For a nondegenerate, correct reference,
the second-order perturbation theory passes the third KL test, whereas for a degenerate and/or incorrect
reference the theory fails the test. 

The third term contains the divergent anomalous contribution in its diagonal summand,
\begin{eqnarray}
- \frac{\beta}{2} \sum_p^{\epsilon_p=\epsilon_h} f_p^- f_p^+ \left( \sum_{r}  \langle pr || pr \rangle  f_r^- \right)^2  ,
\end{eqnarray}
which is essentially the same as the anomalous contribution ``$\Omega_{2A}$'' or Eq.\ (22) of Kohn and Luttinger \cite{kohn}. 
As pointed out by these authors, this divergence is canceled exactly by a term involving $(\mu^{(1)})^2$ 
[Eq.\ (18) of Ref.\ \onlinecite{kohn}] in an isotropic system. In our formalism that is valid for a general system, the whole diagonal sum in the third term is canceled exactly by the penultimate term involving $(\mu^{(1)})^2$, i.e.,
\begin{eqnarray}
&& -\frac{\beta}{2} \sum_{p}^{\epsilon_p=\epsilon_h} {| F_{pp} |^2 f_p^- f_p^+}
+ \frac{\beta}{2} \left(\frac{\sum_p^{\epsilon_p=\epsilon_h} F_{pp} f_p^- f_p^+}{\sum_p^{\epsilon_p=\epsilon_h} f_p^- f_p^+}\right)^2 \sum_p^{\epsilon_p=\epsilon_h} f_p^- f_p^+
=0, \nonumber\\ 
\end{eqnarray}
which may appear to lend support to the Luttinger--Ward prescription \cite{kohn,luttingerward,balian}.
However, it falls short of fundamentally addressing the Kohn--Luttinger conundrum because the 
fourth term of Eq.\ (\ref{Omega2_reduced_limit_deg}) still persists at $T=0$ and it diverges if the degeneracy is lifted at the first order of HCPT.

\begin{table} 
\caption{\label{tab:deg_Omega} Comparison of the zeroth-, first-, and second-order corrections to the grand potential ($\Omega^{(n)}$, $0 \leq n \leq 2$)
as a function of temperature ($T$) for the square-planar H$_4$ molecule (0.8~\AA) in the STO-3G basis set.}
{
\begin{ruledtabular}
\begin{tabular}{lddd}
{$T /~\text{K}$} 
&\multicolumn{1}{c}{$\Omega^{(0)} / E_{\text{h}}$} 
&\multicolumn{1}{c}{$\Omega^{(1)} / E_{\text{h}}$} 
&\multicolumn{1}{c}{$\Omega^{(2)} / E_{\text{h}}$} \\ \hline
0 (HCPT)\footnotemark[1] & 1.7886 & -3.7015 & -0.0222 \\
0 (HCPT)\footnotemark[2] & 1.7886 & -3.6696 & -0.0569 \\
0 (MPPT)\footnotemark[3] & 1.7886 & -3.5817 & \multicolumn{1}{c}{$-\infty$} \\
0 (HF)\footnotemark[4] & 1.7886 & -3.3771 & \multicolumn{1}{c}{$\cdots$} \\
$10^2$ & 1.7877 & -3.3771 & -171.9934 \\
$10^3$ & 1.7798 & -3.3771 & -17.2244 \\
$10^4$ & 1.7008 & -3.3771 & -1.7476 \\
$10^5$ & 0.7938 & -3.3698 & -0.3573 \\
$10^6$ & -14.1403 & -3.5757 & -0.0881 \\
\end{tabular}
\footnotetext[1]{The correct zero-temperature limit. $E^{(n)}-\mu^{(n)}\bar{N}$ at $T=0$ according to the Hirschfelder--Certain degenerate perturbation theory \cite{Hirschfelder} for the triplet ground state. See the corresponding footnote of Table \ref{tab:deg_U}.}
\footnotetext[2]{$E^{(n)}-\mu^{(n)}\bar{N}$ at $T=0$ according to the Hirschfelder--Certain degenerate perturbation theory \cite{Hirschfelder} for the singlet ground state. See the corresponding footnote of Table \ref{tab:deg_U}.}
\footnotetext[3]{$E^{(n)}-\mu^{(n)}\bar{N}$ at $T=0$ according to the M{\o}ller--Plesset perturbation theory \cite{moller}. See the corresponding footnote of Table \ref{tab:deg_U}.}
\footnotetext[4]{The zero-temperature limit of the finite-temperature Hartree--Fock theory. See the corresponding footnote of Table \ref{tab:deg_U}.}
\end{ruledtabular}
}
\end{table}

Table \ref{tab:deg_Omega} confirms the foregoing conclusions numerically for the square-planar H$_4$. 
The correct zero-temperature limits are given in the first row of the table. The zeroth-order grand potential $\Omega^{(0)}$ approaches $E_0^{(0)} = 1.7886\,E_\text{h}$ as $T \to 0$, although the convergence is much slower than $U^{(0)}$, which may be due to the entropy
term in the former. The first-order grand potential $\Omega^{(1)}$ converges at the wrong zero-temperature limit of $-3.3771\,E_\text{h}$, which is higher than
the correct limit of $-3.7015\,E_\text{h}$. The second-order grand potential 
$\Omega^{(2)}$ shows a clear sign of divergence as $T \to 0$. 

\section{Derivations of $\Omega^{(n)}$, $U^{(n)}$, and $\mu^{(n)}$ ($0 \leq n \leq 2$) \label{app:derivation}}

The SoS and reduced analytical formulas for  $\Omega^{(n)}$, $U^{(n)}$, and $\mu^{(n)}$ ($0 \leq n \leq 2$)  in the grand
canonical ensemble are derived succinctly here. A reader is referred to Refs.\ \onlinecite{HirataJha,HirataJha2} for a complete derivation.

The grand partition function $\Xi$ is defined by
\begin{eqnarray}
\Xi = \sum_I e^{- \beta E_I + \beta \mu N_I} , \label{Xidef}
\end{eqnarray}
where $E_I$ and $N_I$ are the FCI energy and number of electrons in the $I$th state, and the summation
runs over all states with any number of electrons (including zero) spanned by a finite basis set.
The chemical potential $\mu$ is determined by the condition \cite{JhaHirata},
\begin{eqnarray}
\bar{N} &=& \frac{1}{\beta} \frac{\partial}{\partial \mu}\ln \Xi \label{Nbar0} \\
&=& \frac{ \sum_I N_I e^{- \beta E_I + \beta \mu N_I} }{\sum_I e^{- \beta E_I + \beta \mu N_I}  }, \label{Nbar}
\end{eqnarray}
where $\bar{N}$ is the correct average number of electrons that keeps the system electrically neutral.
The grand potential $\Omega$ and internal energy $U$ are related to $\Xi$ by 
\begin{eqnarray}
\Omega &=& -\frac{1}{\beta} \ln \Xi, \label{Omegadef}\\
U &=& - \frac{\partial}{\partial \beta} \ln \Xi + \mu \bar{N} \label{Udef1},
\end{eqnarray}
the latter being equivalent to Eq.\ (\ref{U_FCI}).

The $n$th-order perturbation correction to quantity $X$ is defined by
\begin{eqnarray}
X^{(n)} = \left.\frac{1}{n!} \frac{\partial^n X(\lambda)}{\partial \lambda^n}\right|_{\lambda=0}.  \label{X_lambda}
\end{eqnarray}
Here, $X$ can be $\Xi$, $\Omega$, $U$, $\mu$, or $E_I$.

Differentiating both sides of Eq.\ (\ref{Omegadef}) with respect to $\lambda$, 
we readily obtain the SoS formulas for $\Omega^{(n)}$ as
\begin{eqnarray}
\Omega^{(0)} &=& -\frac{1}{\beta} \ln \sum_I e^{- \beta E_I^{(0)} + \beta \mu^{(0)} N_I}, \label{Omega0_SoS}\\
\Omega^{(1)} &=& \langle E_I^{(1)}  - \mu^{(1)} N_I \rangle, \label{Omega1_SoS} \\
\Omega^{(2)} &=& \langle E_I^{(2)}  - \mu^{(2)} N_I \rangle -\frac{\beta}{2} \langle (E_I^{(1)} - \mu^{(1)}N_I)^2 \rangle \nonumber\\
&& + \frac{\beta}{2} \langle E_I^{(1)}- \mu^{(1)}N_I \rangle^2 ,  \label{Omega2_SoS} 
\end{eqnarray}
where $\langle X_I \rangle$ is the zeroth-order thermal average defined by Eq.\ (\ref{X}),
and $E^{(n)}_I$ is identified as the $n$th-order HCPT energy correction \cite{Hirschfelder} for the $I$th state. 

The $\lambda$-differentiation of Eq.\ (\ref{U_FCI}) leads to Eqs.\ (\ref{U0_SoS})--(\ref{U2_SoS}) as the SoS formulas
for $U^{(n)}$.

Likewise, differentiating Eq.\ (\ref{Nbar}), we arrive at the SoS formulas for $\mu^{(n)}$, which read
\begin{eqnarray}
\bar{N} &=& \langle N_I \rangle, \label{mu0_SoS} \\
\mu^{(1)} &=& \frac{\langle E_I^{(1)} (N_I - \bar{N})\rangle }{\langle N_I (N_I - \bar{N}) \rangle }, \label{mu1_SoS} \\
\mu^{(2)} &=& \frac{\langle E^{(2)}_I  (N_I - \bar{N})\rangle }{\langle N_I(N_I - \bar{N}) \rangle} 
- \frac{\beta}{2} \frac{\langle ( E^{(1)}_I - \mu^{(1)} N_I )^2 (N_I - \bar{N})\rangle }{\langle N_I(N_I - \bar{N}) \rangle} .
\nonumber\\ \label{mu2_SoS} 
\end{eqnarray}

These SoS formulas can be reduced to the sum-over-orbitals expressions by combining 
the Boltzmann-sum identities listed in Appendix A of Ref.\ \onlinecite{HirataJha2} 
with the sum rules of the HCPT energy corrections such as
\begin{eqnarray}
\sum_I^{\text{degen.}} E_I^{(1)} &=& \sum_I^{\text{degen.}} \left\{ \sum_{i}^I H_{ii}^{\text{core}} + \sum_{i < j}^I \langle ij || ij \rangle - \sum_i^I \epsilon_i \right\} , \label{E1sumrule} \\
\sum_I^{\text{degen.}} E_I^{(2)} &=& \sum_I^{\text{degen.}}\left\{  \sum_{i,a}^{I,\,\text{denom.}\neq 0} \frac{ \left |H^{\text{core}}_{ia} + \sum_j^I \langle ij||aj \rangle \right |^2 }{\epsilon_i - \epsilon_a} \right. \nonumber\\ 
&& \left. + \sum_{i < j,a < b}^{I,\,\text{denom.}\neq 0} \frac{|\langle ij||ab\rangle|^2}{\epsilon_i + \epsilon_j - \epsilon_a - \epsilon_b} \right\} , \label{E2sumrule}
\end{eqnarray}
where ``degen.''\ means that $I$ runs over all Slater determinants in the degenerate subspace, 
and ``$I,\,\text{denom.}\neq 0$'' excludes summands with a vanishing denominator. 
These sum rules, discussed in detail in Appendix B of Ref.\ \onlinecite{HirataJha2}, are derived by applying
the Slater--Condon rules to the HCPT energy correction formulas \cite{Hirschfelder} and using 
the trace invariance.

This process converts Eqs.\ (\ref{Omega0_SoS}), (\ref{Omega1_SoS}), and (\ref{Omega2_SoS}) into
Eqs.\ (\ref{Omega0_reduced}), (\ref{Omega1_reduced}), and (\ref{Omega2_reduced}), respectively, after
 tedious, but  straightforward algebraic transformations, which are described in detail
in Refs.\ \onlinecite{HirataJha,HirataJha2}. 

Similarly, the reduced formulas for $U^{(0)}$ [Eq.\ (\ref{U0_reduced})], $U^{(1)}$ [Eq.\ (\ref{U1_reduced})], $\mu^{(0)}$
[Eq.\ (\ref{mu0_reduced})], and $\mu^{(1)}$ [Eq.\ (\ref{mu1_reduced})] are derivable by this method \cite{HirataJha}. 
However, a more expedient way is to start with the following identities:
\begin{eqnarray}
U^{(1)} &=& \Omega^{(1)} + \mu^{(1)}\bar{N} + \beta \left( \frac{\partial \Omega^{(1)}}{\partial \beta} \right)_{\mu^{(0)},\,\mu^{(1)}}, \label{U2fromOmega1} \\
U^{(2)} &=& \Omega^{(2)} + \mu^{(2)}\bar{N} + \beta \left( \frac{\partial \Omega^{(2)}}{\partial \beta} \right)_{\mu^{(0)},\,\mu^{(1)},\,\mu^{(2)}}, \label{U2fromOmega2}
\end{eqnarray} 
and
\begin{eqnarray}
\left(\frac{\partial \Omega^{(1)}}{\partial \mu^{(0)}}\right)_{\mu^{(1)}}  = 
\left(\frac{\partial \Omega^{(2)}}{\partial \mu^{(0)}}\right)_{\mu^{(1)},\,\mu^{(2)}}  = 0 , \label{Omega2mu0} 
\end{eqnarray} 
whose justifications are given in Ref.\ \onlinecite{HirataJha2}. 
Substituting Eq.\ (\ref{Omega1_reduced}) into these, we can immediately recover Eq.\ (\ref{U1_reduced})
for $U^{(1)}$ and Eq.\ (\ref{mu1_reduced}) for $\mu^{(1)}$. 
Starting with Eq.\ (\ref{Omega2_reduced}), we arrive at Eq.\ (\ref{U2_reduced})
for $U^{(2)}$ and Eq.\ (\ref{mu2_reduced}) for $\mu^{(2)}$. 

The SoS analytical formulas for $F^{(n)}$ and $U^{(n)}$ ($0 \leq n \leq 3$) in the canonical ensemble can be derived analogously \cite{JhaHirata_canonical}. They do not seem to lend themselves to a reduction to sum-over-orbitals formulas.

%
\end{document}